\documentclass[twocolumn,showpacs,nofootinbib,preprintnumbers,amsmath,amssymb, floatfix]{revtex4}

\usepackage{graphicx}
\usepackage{dcolumn}
\usepackage{bm}

\def\bear{\begin{eqnarray}}
\def\ear{\end{eqnarray}}

\begin{document}

\title{Schwinger Pair Production in Pulsed Electric Fields}

\author{Sang Pyo Kim}\email{sangkim@kunsan.ac.kr}
\affiliation{Department of Physics, Kunsan National University, Kunsan 573-701, Korea}
\affiliation{International Center for Relativistic Astrophysics Network, Piazza della Repubblica, 10
          65122 Pescara, Italy}

\author{Hyung Won Lee}\email{hwlee@inje.ac.kr}
\affiliation{Institute of Basic Science and Department of Computer Aided Science,
          Inje University, 197 Inje-ro, Gimhae 621-749, Korea}
\affiliation{International Center for Relativistic Astrophysics Network, Piazza della Repubblica, 10
          65122 Pescara, Italy}

\author{Remo Ruffini}\email{ruffini@icra.it}
\affiliation{Dipartimento di Fisica and International Center for Relativistic Astrophysics Network,
Univrsita di Roma ``La Sapienza,'' Piazzale Aldo Moro 5, I-00185 Roma, Italy}
\affiliation{International Center for Relativistic Astrophysics Network, Piazza della Repubblica, 10
          65122 Pescara, Italy}

\medskip

\date{\today}

\begin{abstract}
We numerically investigate the temporal behavior and the structure of longitudinal momentum spectrum and the field polarity effect on
pair production in pulsed electric fields in scalar quantum electrodynamics (QED). Using the evolution operator
expressed in terms of the particle and antiparticle operators, we find
the exact quantum states under the influence of electric pulses
and measure the number of pairs of the Minkowski particle and antiparticle.
The number of pairs, depending on the configuration of electric pulses, exhibits rich structures
in the longitudinal momentum spectrum and undergoes diverse dynamical behaviors at the onset of the interaction 
but always either converges to a momentum-dependent constant or oscillates around a momentum-dependent 
time average after the completion of fields.

\end{abstract}

\maketitle

\section{Introduction}

Schwinger pair production by temporally or spatially inhomogeneous electric fields has been
a theoretically and experimentally interesting and challenging issue.
The Minkowski or Dirac vacuum under the influence of a strong electric field becomes unstable
and emits electron-positron or charged particle-antiparticle pairs.
The possibility of directly measuring electron-positron pairs by strong laser sources such as Extreme Light
Infrastructure (ELI) has recently boosted intensive researches on pair production by pulsed electric fields
\cite{Ringwald01,AHRSV01,BPRSS06,BMNNP10} (for review and references, see also Refs.\,\cite{Dunne09,BPRRSSS09,DMHK11}).
In contrast to vacuum polarization, pair production has been studied for various
electromagnetic configurations, for which many analytical approximation schemes
have been elaborated \cite{Dunne09}.

The Keldysh approach \cite{BrezinItzykson70,Popov72} (review and references, see
Ref. \cite{Popov04}), the worldline instanton method \cite{DunneSchubert05,DWGS06} and the phase-integral
or WKB method \cite{KimPage02,KimPage07,KRX08,DumluDunne10,DumluDunne11,DumluDunne11b}, to list a few,
have been widely used as analytical approximation schemes.
The two typical methods employed to numerically compute the pair-production rate in time-dependent electric fields
are the kinetic approach \cite{SGD08,HAG08,HADG09,Dumlu09,DLMH09,DGS09,Dumlu10} and the Wigner formalism \cite{HAG10,OHA11,HAG11}.
The quantum Vlasov equation has been often used for numerical calculation of pair production
with or without back-reaction included \cite{KESCM91,KESCM92,SRSBTP97,KME98,SBRSPT98,PSRSP05,FGKS11,KimSchubert11}.

The main purpose of this paper is to employ the evolution operator formalism
for numerical computation of pair production by pulsed electric fields of various configurations
in scalar QED. The stratagem is to express the Hamiltonian
in terms of the creation and annihilation operators of the Minkowski vacuum
and then find the evolution operator satisfying the time-dependent Schr\"{o}dinger equation.
In fact, the Hamiltonian for a spinless charged boson in a time-dependent
electric field in the momentum space consists of infinite
number of oscillators with time-dependent frequencies and
each oscillator can have the evolution operator represented
in $SU(1,1)$ algebra.
We then choose the number operator, the one-pair creation
operator and the one-pair annihilation operator in the Minkowski vacuum
as the generators for the evolution operator
\begin{eqnarray}
\prod_k e^{\xi_k (t) \hat{a}^{\dagger}_k \hat{b}^{\dagger}_{-k}}
e^{i \gamma_k (t) (\hat{a}^{\dagger}_k \hat{a}_k + \hat{b}_{-k} \hat{b}^{\dagger}_{-k})}
e^{\eta_k (t) \hat{a}_k \hat{b}_{-k}}. \nonumber
\end{eqnarray}

In quantum optics the evolution operator for a time-dependent oscillator has been studied
in a different representation of $SU(1,1)$ \cite{Lo91}.
The oscillator representation or the creation and annihilation operator representation
has also been used to derive the quantum Vlasov equation
for pair production \cite{BalantekinFricke91,BSF91,SBRSPT98,PSRSP05,FGKS11,KimSchubert11}.
The difference of this paper from other earlier works is that we
find the evolution operator by directly solving the time-dependent
Schr\"{o}dinger equation, then find the exact quantum state evolved
from the Minkowski vacuum, and finally compute the number of pairs
during the whole evolution in pulsed electric fields.
The evolution operator represented by the creation and annihilation operators
has an additional advantage of expressing the exact quantum state as
the squeezed vacuum of multi-pairs of particle and antiparticle
under the influence of the pulsed electric fields.
Further it provides a good measure for counting the number of Minkowski particle and
antiparticle pairs.

To study pair production from nontrivial configurations of
electromagnetic field, we shall consider only pulsed electric fields, simplified model fields,
which act for a finite duration but is uniform over the space.
Thus the Minkowski vacuum excites into multi-particle and antiparticle state during
the interaction of the electric field and finally settles down in some nontrivial vacuum state and,
in an exceptional case, returns to the Minkowski vacuum.
The pulsed electric fields to be studied are classified into
the mono-polarity type and the di-polarity type.
In the first class we consider (i) $E(t) = E_0 /\cosh^2 (t/\tau)$,
(ii) $E(t) = E_0 e^{- t^2/\tau^2 }$,
(iii) $A_{\parallel} (t) = (E_0/\omega) e^{- t^2/\tau^2} \cos(\omega t)$,
and in the second class we also consider
(iv) $E(t) = E_0 /(\cosh^2 ((t-t_1)/\tau) - E_0 /\cosh^2 ((t-t_2)/\tau)$,
(v) $A_{\parallel} (t) = E_0 \tau /(1 + t^2 / \tau^2)$,
(vi) $A_{\parallel} (t) = \sqrt{n(n+1)} \tau/\cosh(t/\tau)$.

The electric fields (i)-(vi) provides a good arena to discuss some fundamental questions,
not to mention possible applications to ultra-strong lasers. The electric fields
(i)-(iii) of mono-polarity type necessarily lead to non-zero gauge potential values,
in which the adiabatic basis differs from the Minkowski one. Then a question may be raised
whether pairs produced by electric pulses are the Minkowski ones or the adiabatic ones 
\cite{KME98,KimSchubert11}.
And the oscillation of the number of pairs after the completion of electric pulses
may be another interesting question since the kinetic approach using the adiabatic basis predicts
non-oscillating pair production  while the nonadiabatic method predicts
oscillating pair production for general electric fields 
except for the solitonic gauge field \cite{KimSchubert11}.
The evolution operator provides asymptotic solutions in the remote future for all pulsed electric fields
including (i)-(vi), according to which the number of pairs converges to a momentum-dependent constant
when the gauge potentials vanish or oscillates around a momentum-dependent time-averaged constant
when the gauge potentials have nonzero constant.

Another interesting issue is the structure of the longitudinal momentum spectrum of
pairs discovered by Hebenstreit et al in a sinusoidal field with Gaussian envelope \cite{HADG09}.  
Dumlu and Dunne explained the substructure of the spectrum in the inverse square gauge potential 
by the Stokes phenomenon, in which more than two pairs of complex turning points for the Fourier mode equation
contribute either in phase or out of phase to the WKB instanton action \cite{DumluDunne10,DumluDunne11}.
The analytical approximation schemes such as the Keldysh approach or the WKB or phase-integral method without
Stokes phenomenon cannot explain the substructure of the longitudinal momentum distribution.  
The evolution operator formalism is efficient enough to compute the longitudinal momentum distribution
of produced pairs in the pulsed electric fields (i)-(vi) and confirms the substructure of the spectrum
for the critical field strength and Compton time scale. The spectra for (i), (ii), (iv)
and (v) exhibit rich structures but those for (iii) and (vi) have relatively simple structure.
Bunching of spectrum in the field (iii) is observed around one positive and one negative momentum with the same 
magnitude. Further the polarity of the electric field (iv) leads
to negative momentum dominance or positive momentum dominance in the spectrum.

The organization of this paper is as follows. In Sec. II, we introduce the evolution operator
in algebraic form for scalar QED in a pulsed electric field and express the vacuum polarization
in terms of the complex parameters for the evolution operator. And we express
the number of pairs by the imaginary part of the parameter for the number operator.
In Sec. III, by solving the time-dependent Schr\"{o}dinger equation,
we derive a set of first order differential equations for three complex parameters
for the evolution operator. And we find the asymptotic solutions for the pulsed electric fields
whose gauge potentials approach either zero or nonzero constant
after the completion of the interaction.
In Sec. IV we give an intuitive interpretation of pair production by pulsed electric fields
from the analogy with quantum mechanics.
In Sec. V, we numerically investigate pair production by several configurations of electric fields,
some of which have recently been studied in literature.
In Sec. VI, we discuss the physical implications of the results and conclude the paper.

\section{Evolution Operator Approach}

In scalar QED the Fourier-decomposed Hamiltonian for a spinless charged boson with mass $m$ in an electric field along a fixed direction
\cite{KimSchubert11}
\begin{eqnarray}
\hat{H} (t) &=& \int \frac{d^3 k}{(2 \pi)^3} \Bigl[\hat{\pi}_k^{\dagger} \hat{\pi}_k
+ \omega^2_k (t) \hat{\phi}_k^{\dagger} \hat{\phi}_k \Bigr],
\nonumber\\ \omega_k^2 (t) &=& (k_{\parallel} - qA_{\parallel} (t))^2 + {\bf k}_{\perp}^2 + m^2, \label{f-ham}
\end{eqnarray}
may have an oscillator representation
\begin{eqnarray}
\hat{H}_k (t) = \frac{1}{2}\Omega^{(+)}_k (t) \hat{N}_k + \frac{1}{2} \Omega^{(-)}_k (t) (\hat{J}^{(+)}_k + \hat{J}^{(-)}_k ),
\end{eqnarray}
where
\begin{eqnarray}
\Omega_k^{(\pm)} (t) = \frac{\omega^2_k (t) \pm \omega^2_k (t_0)}{\omega_k (t_0)},
\end{eqnarray}
and
\begin{eqnarray}
\hat{N}_k = \hat{a}^{\dagger}_k \hat{a}_k + \hat{b}_{-k} \hat{b}^{\dagger}_{-k}, ~
\hat{J}^{(+)}_k = \hat{a}^{\dagger}_k \hat{b}^{\dagger}_{-k},~
\hat{J}^{(-)}_k = \hat{a}_k \hat{b}_{-k}.
\end{eqnarray}
Here we have used the oscillator representation in the Minkowski vacuum
\begin{eqnarray}
\hat{\phi}_k &=& \frac{1}{\sqrt{2 \omega_k (t_0)}} (\hat{a}_k + \hat{b}^{\dagger}_{-k}), \nonumber\\
\hat{\pi}_k &=&  i \sqrt{\frac{\omega_k (t_0)}{2}} (\hat{a}^{\dagger}_k -  \hat{b}_{-k}),
 \label{Min-op}
\end{eqnarray}
where $\hat{a}_k$ and $\hat{b}^{\dagger}_{-k}$ are the particle and antiparticle operators.
And these operators constitute the $SU(1,1)$ algebra
\begin{eqnarray}
[ \hat{N}_k, \hat{J}^{(\pm)}_k ] = \pm 2 \hat{J}^{(\pm)}_k, \quad [\hat{J}^{(+)}_k, \hat{J}^{(-)}_k] = \hat{N}_k.
\end{eqnarray}
In this paper we shall focus on pulsed electric fields that act effectively for a finite duration.
Before the onset of a pulsed electric field, we may choose a gauge
\begin{eqnarray}
A_{\parallel} (t) = 0, \quad E(t) = 0 \quad (t \leq t_0),
\end{eqnarray}
so that the in-vacuum is the Minkowski vacuum.
The constant electric field with $A_{\parallel} = - E_0 t$ will not be considered in this paper
since no gauge can be chosen to make the initial vacuum
the Minkowski one, for which the in- and out-vacua are defined as asymptotic states.

The evolution of the Minkowski vacuum follows the time-dependent Schr\"{o}dinger equation
\begin{eqnarray}
i \frac{\partial \hat{U}_k (t)}{\partial t} = \hat{H}_k (t) \hat{U}_k (t). \label{ev eq}
\end{eqnarray}
The evolution operator in time-ordered integral
\begin{eqnarray}
\hat{U}_k (t) = {\rm T} \Bigl( e^{ - i \int_{t_0}^t \hat{H}_k (t') dt' } \Bigr),
\end{eqnarray}
does not provide useful information about the quantum state unless $\Omega_k^{(-)} (t) = 0$ since the
Hamiltonian is then entangled such that $[\hat{H}_k (t''), \hat{H}_k (t')] \neq 0$ for $t'' \neq t'$.
The particle and antiparticle operators evolve as
\begin{eqnarray}
\hat{a}_k (t) = \hat{U}_k (t) \hat{a}_k \hat{U}^{\dagger}_k (t) := \mu_k (t) \hat{a}_k + \nu^*_k \hat{b}^{\dagger}_{-k}, \nonumber\\
\hat{b}_{-k} (t) = \hat{U}_k (t) \hat{b}_{-k} \hat{U}^{\dagger}_k (t) := \mu_k (t) \hat{b}_{-k} + \nu^*_k \hat{a}^{\dagger}_{k}.
\label{bog tran}
\end{eqnarray}
The time-dependent vacuum state is given by
\begin{eqnarray}
\vert 0, t \rangle = \prod_k \hat{U}_k (t) \vert 0, {\rm in} \rangle,
\end{eqnarray}
where $\vert 0, {\rm in} \rangle$ denotes the Minkowski vacuum.
The number of the Minkowski particle carried by the time-dependent vacuum, which is equal to the number of time-dependent particle
carried by the Minkowski vacuum, is
\begin{eqnarray}
\langle 0, t \vert \ \hat{a}^{\dagger}_k \hat{a}_k \vert 0, t \rangle = \langle 0, {\rm in} \vert
\hat{a}^{\dagger}_k (t) \hat{a}_k (t) \vert 0, {\rm in} \rangle = |\nu_k|^2.
\end{eqnarray}
The same is true for the antiparticle production.
Further, the vacuum persistence is related to the mean number \cite{KLY08}
\begin{eqnarray}
\vert \langle 0, t \vert 0, {\rm in} \rangle \vert^2 = \exp \Bigl[- \sum_k \ln (1 + |\nu_k|^2)  \Bigr]
\end{eqnarray}

On the other hand, the $SU(1,1)$ algebra may lead to the evolution operator of the form
\begin{eqnarray}
\hat{U}_k (t) = e^{\xi_k (t) \hat{J}^{(+)}_k} e^{i \gamma_k (t) \hat{N}_k} e^{\eta_k (t) \hat{J}^{(-)}_k}. \label{ev op}
\end{eqnarray}
Here the complex parameters $\xi_k$, $\gamma_k$ and $\eta_k$ will be determined from the evolution equation
(\ref{ev eq}).
It should be mentioned that Eq.\,(\ref{ev op}) differs from Eq.\,(31) of Ref. \cite{Lo91}
for a time-dependent oscillator, which would read
\begin{eqnarray}
\hat{J}_k^{(+)} &=& i \hat{\phi}_k^{\dagger}
\hat{\phi}_k, \quad \hat{J}_k^{(-)} = i \hat{\pi}_k^{\dagger} \hat{\pi}_k, \nonumber\\
\hat{J}_k^{(0)} &=& \frac{i}{4} (\hat{\pi}_k \hat{\phi}_k + \hat{\phi}_k \hat{\pi}_k + \hat{\pi}^{\dagger}_k
\hat{\phi}^{\dagger}_k + \hat{\phi}^{\dagger}_k \hat{\pi}^{\dagger}_k).
\end{eqnarray}
Another form of the evolution operator is realized by inverting the Bogoliubov transformation (\ref{bog tran}) in Ref.\,\cite{KLY08}.
The representation (\ref{ev eq}) is particularly useful for pair production
since it expresses the out-vacuum as a squeezed vacuum of the Minkowski vacuum
\begin{eqnarray}
\vert 0, t \rangle &=& \prod_k e^{i \gamma_k (t)}  e^{\xi_k (t) \hat{J}^{(+)}_k}  \vert 0, {\rm in} \rangle \nonumber\\
&=& \prod_k e^{i \gamma_k (t)} \Bigl( \sum_{n = 0}^{\infty} \xi^n_k (t)  \vert n_k, \bar{n}_{-k}, {\rm in} \rangle \Bigr),
\label{sq vac}
\end{eqnarray}
where $n_k$ and $\bar{n}_{-k}$ denote the $n$ particles with momentum $k$ and $n$ antiparticles with momentum $-k$,
respectively. Thus pair production conserves charge and momentum.
It further leads to the vacuum polarization
\begin{eqnarray}
\langle 0, t \vert 0, {\rm in} \rangle = \exp \Bigl[- i \sum_k \gamma_k^* (t) \Bigr],
\end{eqnarray}
and the vacuum persistence
\begin{eqnarray}
\vert \langle 0, t \vert 0, {\rm in} \rangle \vert^2 = \exp \Bigl[- 2 \sum_k \gamma_{{\rm I} k} (t)  \Bigr],
\end{eqnarray}
where $\gamma_{{\rm I} k}$ denotes the imaginary part of $\gamma_k$.
Therefore we can find the number of produced pairs from the imaginary part of the parameter $\gamma_k (t)$
during the interaction of the electric field as well as after the completion of the interaction
\begin{eqnarray}
|\nu_k (t)|^2 = e^{2 \gamma_{{\rm I} k} (t)} - 1.
\end{eqnarray}

\section{Master Equations for Evolution Operator}

The time-dependent Schr\"{o}dinger equation (\ref{ev eq}) together with $SU(1,1)$ algebra leads to the differential equations for
the complex parameters \footnote{Here we have used the algebraic relation, $e^{\lambda \hat{N} } \hat{M} e^{-\lambda \hat{N}} =
\sum_{n = 0}^{\infty} \frac{\lambda^n}{n!} [\hat{N}, \hat{M}]^{(n)}$, where $[\hat{N}, \hat{M}]^{(0)} = \hat{M}$ and
$[\hat{N}, \hat{M}]^{(n)} = [\hat{N}, [\hat{N}, \hat{M}]^{(n-1)}]$.}
\begin{eqnarray}
\dot{\gamma}_k + i \dot{\eta}_k e^{-2 i \gamma_k} \xi_k = - \frac{1}{2} \Omega^{(+)}_k, \\
\dot{\xi}_k - 2 i \dot{\gamma}_k \xi_k  + \dot{\eta}_k e^{-2 i \gamma_k} \xi^2_k = - \frac{i}{2} \Omega^{(-)}_k, \\
\dot{\eta}_k e^{-2 i \gamma_k} = - \frac{i}{2} \Omega^{(-)}_k.
\end{eqnarray}
The differential equations for the real and imaginary parts of complex parameters can be grouped into the set
that determines pair production
\begin{eqnarray}
\dot{\gamma}_{{\rm I} k} &=& - \frac{1}{2} \Omega^{(-)}_k  \xi_{{\rm I} k}, \label{gamma-I}\\
\dot{\xi}_{{\rm R} k} &=& \Omega^{(+)}_k \xi_{{\rm I} k} + \Omega^{(-)}_k \xi_{{\rm R} k} \xi_{{\rm I} k}, \label{xi-R}\\
\dot{\xi}_{{\rm I} k} &=& - \Omega^{(+)}_k \xi_{{\rm R} k} - \frac{1}{2} \Omega^{(-)}_k (1 + \xi^2_{{\rm R} k} -\xi^2_{{\rm I} k}),
\label{xi-I}
\end{eqnarray}
and into another set that are relevant for the vacuum polarization
\begin{eqnarray}
\dot{\gamma}_{{\rm R} k} &=& - \frac{1}{2} \Omega^{(+)}_k - \frac{1}{2} \Omega^{(-)}_k \xi_{{\rm R} k}, \\
\dot{\eta}_{{\rm R} k} &=& \frac{1}{2} \Omega^{(-)}_k e^{-2 \gamma_{{\rm I} k}} \sin (\gamma_{{\rm R} k}), \\
\dot{\eta}_{{\rm I} k} &=& - \frac{1}{2} \Omega^{(-)}_k e^{-2 \gamma_{{\rm I} k}} \cos (\gamma_{{\rm R} k}).
\end{eqnarray}

To get the initial data for the parameters, we employ the Baker-Campbell-Hausdorff formula to write the evolution
in a single exponential operator
\begin{eqnarray}
\hat{U}_k (t) = \exp \Bigl [  (1 - i \gamma_k (t)) (\xi_k (t) \hat{J}^{(+)}_k + \eta_k (t) \hat{J}^{(-)}_k ) \nonumber\\
+ \Bigl(i \gamma_k (t) + \frac{1}{2} \xi_k (t) \eta_k (t) (1- i \gamma_k (t) ) \Bigr) \hat{N}_k +  \cdots \Bigr],
\end{eqnarray}
where the dots denote polynomials of $\xi_k$, $\gamma_k$ and $\eta_k$ higher than third order.
Before the onset of the electric field $\Omega^{(-)}_k = 0$, the evolution operator takes the form
\begin{eqnarray}
\hat{U}_k (t) = e^{- \frac{i}{2} \Omega_k (t_0) \hat{N}_k (t-t_0)}.
\end{eqnarray}
Therefore the initial data are given by
\begin{eqnarray}
\xi (t_0) = \gamma_k (t_0) = \eta (t_0) = 0, \quad \dot{\gamma}_k (t_0) = - \frac{1}{2} \Omega_k (t_0).
\end{eqnarray}

For pulsed electric fields  such that $E(t) = 0$ and $A_{\parallel} (t) = {\rm constant}$ for large $t$, we find
the asymptotic solutions. In the first case of $A_{\parallel} (\infty ) = 0$ so that $\Omega^{(-)}_k (\infty) = 0$ and $\Omega^{(+)}_k (\infty) =
2 \omega_k (t_0) = 2 \omega_k (\infty)$, the asymptotic solutions are
\begin{eqnarray}
\gamma_{{\rm I} k} (t) &=& \gamma_{{\rm I} k} (\infty),\nonumber\\
\xi_{{\rm R} k} (t) &=& |\xi_k (\infty)| \cos(2 \omega_k (\infty) t + \varphi_k), \nonumber\\
\xi_{{\rm I} k} (t) &=& - |\xi_k (\infty)| \sin(2 \omega_k (\infty) t + \varphi_k). \label{asym sol1}
\end{eqnarray}
Here three integration constants are $\varphi_k$ and the remaining two are identified
with those at $t = \infty$. This implies that the number of pairs per unit volume
and per unit time approaches constant and the squeezing parameter rotates as $\xi_k = |\xi_k (\infty)|
e^{- i (2 \omega_k (\infty) t + \varphi_k)}$. In the second case of more general
$A_{\parallel} (\infty ) \neq 0$ and $\Omega^{(-)} (\infty ) \neq 0$, the asymptotic solutions are given by
\begin{eqnarray}
\gamma_{{\rm I} k} (t) &=& c_1 - \frac{1}{2} \Omega_k^{(-)} (\infty ) \int^t  \xi_{{\rm I} k}  dt',\nonumber\\
\xi_{{\rm I} k} (t) &=& - \frac{ 2 \omega_k (\infty) }{ \Omega_k^{(-)} (\infty ) }
\frac{\sqrt{1+c_3^2} \cos (2 \omega_k (\infty) t + \vartheta_k) }{c_2 + \sqrt{1+c_3^2} \sin (2 \omega_k (\infty) t + \vartheta_k) }, \nonumber\\
\xi_{{\rm R} k} (t) &=& - \frac{\Omega_k^{(+)}(\infty )}{\Omega_k^{(-)}(\infty )} \Bigl[1 - \Bigl( 1 - 2\frac{\Omega_k^{(-)}(\infty )}{(\Omega_k^{(+)}(\infty ))^2} \dot{\xi}_{{\rm I}k} \nonumber\\
&& - \frac{(\Omega_k^{(-)}(\infty ))^2}{(\Omega_k^{(+)} (\infty ))^2} (1- (\xi_{{\rm I}k})^2) \Bigr)^{1/2} \Bigr],
\label{asym sol2}
\end{eqnarray}
where $c_1$, $c_2$ and $c_3$ are integration constants and $\tan \vartheta_k = 1/c_3$. Further integrating Eq.\,(\ref{asym sol2})
\begin{eqnarray}
\gamma_{{\rm I} k} (t) = c_1 + \frac{1}{2} \ln \Bigl|c_2 + \sqrt{1+c_3^2} \sin (2 \omega_k (\infty) t + \vartheta_k)  \Bigr|,
\end{eqnarray}
we find the number of produced pairs
\begin{eqnarray}
N_k (t) = e^{2c_1}  \Bigl|c_2 + \sqrt{1+c_3^2} \sin (2 \omega_k (\infty) t + \vartheta_k)  \Bigr| -1. \label{asym num2}
\end{eqnarray}
The integration constants of the asymptotic solutions (\ref{asym sol1}) or (\ref{asym sol2}) should be determined
by solving the evolution equations (\ref{gamma-I})-(\ref{xi-I}) or may be found from other analytical methods.

\section{Scattering Picture} \label{scat pic}

The general feature of pair production can be understood from the analogy with quantum mechanical scattering problem.
We write the mode equation for the scalar field as
\begin{eqnarray}
- \frac{d^2 \phi_k}{dt^2} -  P^2_{\parallel} (t) \phi_k = (m^2 + {\bf k}^2_{\perp}) \phi_k, \label{qm eq}
\end{eqnarray}
where the kinematic momentum is
\begin{eqnarray}
P_{\parallel} (t)= k_{\parallel} - qA_{\parallel} (t).
\end{eqnarray}
The mode equation (\ref{qm eq}) in the domain of time can be interpreted as a nonrelativistic particle with mass $1/2$ and
energy $\epsilon = m^2 + {\bf k}_{\perp}^2$ under the potential $V(t) = - P^2_{\parallel} (t)$. We assume that
the pulsed electric field has the gauge potential $A_{\parallel} (- \infty) = 0$ but $A_{\parallel} (\infty) = {\rm constant}$.
Depending on the profile of $A_{\parallel} (t)$ during the interaction
of the electric field pulse, the potential becomes a barrier or well and the asymptotic value $V(\infty)$
may be the same as $V(- \infty)$ or higher (the green line) or lower (the sky blue line) (see Fig. 1).
The incident wave from the future is partially reflected by the well or barrier back into the future and
partially transmitted into the past
\begin{eqnarray}
\alpha_k v_k (t) + \beta_k v_k^* (t)   \longleftrightarrow  u_k (t) ,  \label{scat sol}
\end{eqnarray}
where
\begin{eqnarray}
u_k (t) = \frac{e^{-i \omega_k (- \infty) t}}{\sqrt{2 \omega_k (- \infty)}},
\quad v_k(t) = \frac{e^{-i \omega_k (\infty) t}}{\sqrt{2 \omega_k (\infty)}}.
\end{eqnarray}
First, in the limit of weak field $(|qA_{\parallel}| \ll m)$ or large momentum such that
\begin{eqnarray}
\Bigl| \frac{\omega_k (t) - \omega_k (- \infty)}{\omega_k (- \infty)} \Bigr| \ll 1,
\end{eqnarray}
Eq.\,(\ref{qm eq}) is the scattering by a high energy particle over a shallow well or barrier. Thus the reflection coefficient
$\beta_k$ is suppressed for all $k_{\parallel}$.
Second, in the limit of strong field $(|qA_{\parallel}| \gg m)$ or small momentum such that
\begin{eqnarray}
\Bigl| \frac{\omega_k (t) - \omega_k (- \infty)}{\omega_k (- \infty)} \Bigr| \geq 1,
\end{eqnarray}
Eq.\,(\ref{qm eq}) is the scattering over a relatively deep well or high barrier. The kinetic energy
for small $k_{\parallel}$ is smaller than for large $k_{\parallel}$, which means that
the reflection coefficient for small $k_{\parallel}$ is larger than that for large $k_{\parallel}$.
\begin{figure}[ht]
\includegraphics[width=7cm,height=5cm]{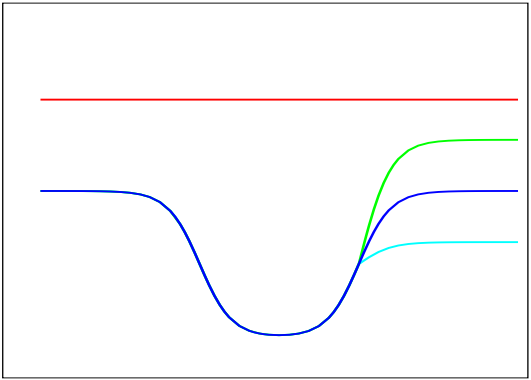}
\caption{(color online). The schematic plot for the scattering problem: the red line denotes the energy of the particle,
$\mu^2 = m^2 + {\bf k}_{\perp}^2$; the blue line denotes a potential, $V(t) = - P_{\parallel}^2 (t))$, where
$A_{\parallel} (\mp \infty) = 0$; the green or sky blue line denote a potential, $V(t) = - P_{\parallel}^2 (t)$, where
$A_{\parallel} (- \infty) = 0$ but $A_{\parallel} (\infty) \neq 0$.} \label{F1}
\end{figure}

The above quantum mechanical interpretation may have a direct interpretation of pair production
when $A_{\parallel} (\infty) = 0$ so that $\omega_k (\infty) = \omega_k (-\infty)$. The positive frequency solution
in the right hand side of Eq.\,(\ref{scat sol}) splits into a positive solution and a negative solution
with the same frequency in the left hand side. Thus $|\beta_k|^2$ is the mean number of produced pairs by the electric field pulse.
However, when $A_{\parallel} (\infty) \neq 0$
and $\omega_k (\infty) \neq \omega_k (-\infty)$, we may write the solution in the future
by another basis with the initial frequency
\begin{eqnarray}
\phi_k (t) = \mu_k u_k (t) + \nu_k u_k^* (t).  \label{scat sol2}
\end{eqnarray}
Using the Wronskian
\begin{eqnarray}
{\rm Wr} [u_k, u_k^*] = i, \quad {\rm Wr} [v_k, v_k^*] = i,
\end{eqnarray}
we find the relation between two set of coefficients
\begin{eqnarray}
\mu_k &=& - i \Bigl( \alpha_k {\rm Wr} [v_k, u_k^*] + \beta_k {\rm Wr} [v_k^*, u_k^*] \Bigr) ,\nonumber\\
\nu_k &=&  i \Bigl( \alpha_k {\rm Wr} [v_k, u_k] + \beta_k {\rm Wr} [v_k^*, u_k] \Bigr).
\end{eqnarray}
Thus the number of produced pairs measured with respect to the Minkowski vacuum
\begin{eqnarray}
|\mu_k (t)|^2 &=&  \Bigl| \frac{1}{2} \Bigl(\sqrt{\frac{\omega_k (\infty)}{\omega_k (-\infty)}} -
\sqrt{\frac{\omega_k (- \infty)}{\omega_k (\infty)}} \Bigr) \alpha_k
\nonumber\\ &&+ \frac{1}{2} \Bigl(\sqrt{\frac{\omega_k (\infty)}{\omega_k (-\infty)}} +
\sqrt{\frac{\omega_k (- \infty)}{\omega_k (\infty)}}\Bigr) \beta_k e^{2 i \omega_k (\infty) t} \Bigl|^2, \nonumber\\
\end{eqnarray}
oscillates with the frequency $2 \omega_k (\infty)$ unless $\omega_k (\infty) = \omega_k (-\infty)$, which confirms the
the result (\ref{asym num2}) from the asymptotic solution. Note that
$|\nu_k (t)|^2$ counted with respect to the Minkowski vacuum is the
same as $|\beta_k (t)|^2$ counted with respect to the adiabatic vacuum
only when $\omega_k (\infty) = \omega_k (-\infty)$. (For discussion on
pair production in the adiabatic basis in QED, see Ref.\,\cite{KME98}.)

A few remarks are in order. First, the kinetic approach
\begin{eqnarray}
\phi_k (t) &=& \bar{\alpha}_k (t) \frac{e^{ - i \int_{- \infty}^t \omega_k (t') dt'}}{\sqrt{2 \omega_k (t)}}
+ \bar{\beta}_k (t) \frac{e^{ i \int_{- \infty}^t \omega_k (t') dt'}}{\sqrt{2 \omega_k (t)}}, \nonumber\\
\dot{\phi}_k (t) &=& - i \omega_k(t) \nonumber\\
&\times& \Bigl(\bar{\alpha}_k (t) \frac{e^{ - i \int_{- \infty}^t \omega_k (t') dt'}}{\sqrt{2 \omega_k (t)}}
- \bar{\beta}_k (t) \frac{e^{ i \int_{- \infty}^t \omega_k (t') dt'}}{\sqrt{2 \omega_k (t)}} \Bigr) \nonumber\\
\label{kin ap}
\end{eqnarray}
applied to the pulsed electric field results in the asymptotic solutions (\ref{scat sol}) with the coefficients
\begin{eqnarray}
\alpha_k (\infty) &=& \bar{\alpha}_k (\infty) e^{-i \int_{- \infty}^{t_1} \omega_k (t') dt' + i \omega_k (\infty) t_1},
\nonumber\\
\beta_k (\infty) &=& \bar{\beta}_k (\infty) e^{i \int_{- \infty}^{t_1} \omega_k (t') dt' - i \omega_k (\infty) t_1}.
\end{eqnarray}
where $t_1$ is an arbitrary time beyond which $\omega_k (t) = \omega_k (\infty)$. Thus the number of pairs from the kinetic approach
is the same from Eq.\,(\ref{scat sol}), that is, $|\beta_k|^2 = |\bar{\beta}_k|^2$.
Second, some gauge potential may involve an odd function $\bar{A}_{\parallel}(t)$ such that for a fixed constant
$c$
\begin{eqnarray}
P_{\parallel} (t) = k_{\parallel} + c - \bar{A}_{\parallel} (t),
\end{eqnarray}
and $P_{\parallel} (-t)$ for $k_{\parallel} + c $ is the negative of $P_{\parallel} (t)$ for $- (k_{\parallel}+c)$.
This implies that the scattering equation for $-(k_{\parallel} + c)$ is the backward scattering for
$(k_{\parallel} + c)$ in time and has the same reflection and transmission coefficients.
Thus the number of pairs is symmetric in the longitudinal momentum around $k_{\parallel} = -c$ in the remote future.

\section{Pair Production in Mono-Polarity Electric Fields} \label{mono-pol}

For the purpose of numerical calculations, we use the Compton unit $\hbar = c = e =  m = 1$ so that
the time is measured in Compton time, the field strength in the critical strength and the energy
in the rest mass energy of the particle:
\begin{eqnarray}
t_C = \frac{\hbar}{mc^2} = 1, \quad E_c = \frac{m^2 c^3}{\hbar e} = 1, \quad mc^2 = 1.
\end{eqnarray}
We restrict to the zero-transverse momentum, which is equivalent to replacing $m^2 =1$ by $\epsilon_{\perp}^2 = m^2 + {\bf k}_{\perp}^2 = 1$
in all the calculations below. In this unit system $\Omega_k^{(\pm)}$ in Eqs.\,(\ref{gamma-I})-(\ref{xi-I}) read
\begin{eqnarray}
\Omega_k^{(+)} (t) &=& \frac{2+ P^2_{\parallel} (t) + k_{\parallel}^2}{\sqrt{1+ k_{\parallel}^2}}, \nonumber\\
\Omega_k^{(-)} (t) &=& \frac{P^2_{\parallel} (t) - k_{\parallel}^2}{\sqrt{1+ k_{\parallel}^2}}.
\end{eqnarray}
Further we shall consider only the strong field regime since pair production is prominent
for an electric field near or above the critical strength and
dynamical characteristics is manifest at the Compton time scale.

\subsection{$E(t) = \frac{E_0}{\cosh^2 (\frac{t}{\tau})}$} \label{Saut}

The Sauter electric field is a frequently used model in QED which is uniform in space but nontrivial in time.
The Green function and the asymptotic number of pairs of the adiabatic particle and antiparticle
have been known \cite{Nikishov70}. The gauge potential is chosen
\begin{eqnarray}
A_{\parallel} (t) = - E_0 \tau \Bigl(1 + \tanh (\frac{t - 10 \tau}{\tau}) \Bigr),
\end{eqnarray}
such that the initial state is the Minkowski vacuum. The duration of the field is effectively
characterized by $2\tau$. We shift the peak of electric field by $10 \tau$ only for the numerical purpose.
\begin{figure}[ht]
\includegraphics[width=0.75\linewidth]{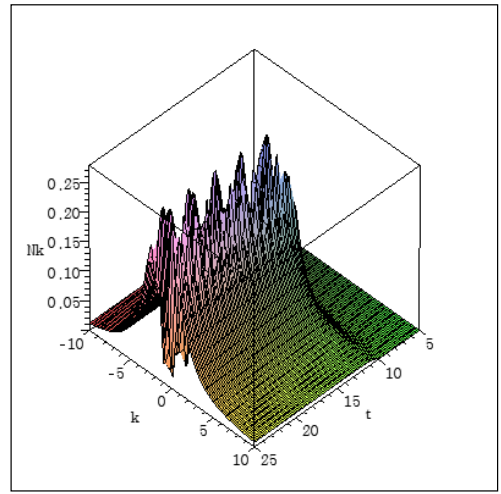}
\caption{(color online). The number of pairs for the Sauter electric field with $E_0 = 1$ and $\tau =1$ is plotted
as a function of time and longitudinal momentum.}
\label{tanh-F1}
\end{figure}
\begin{figure}[t]
{\includegraphics[width=0.475\linewidth]{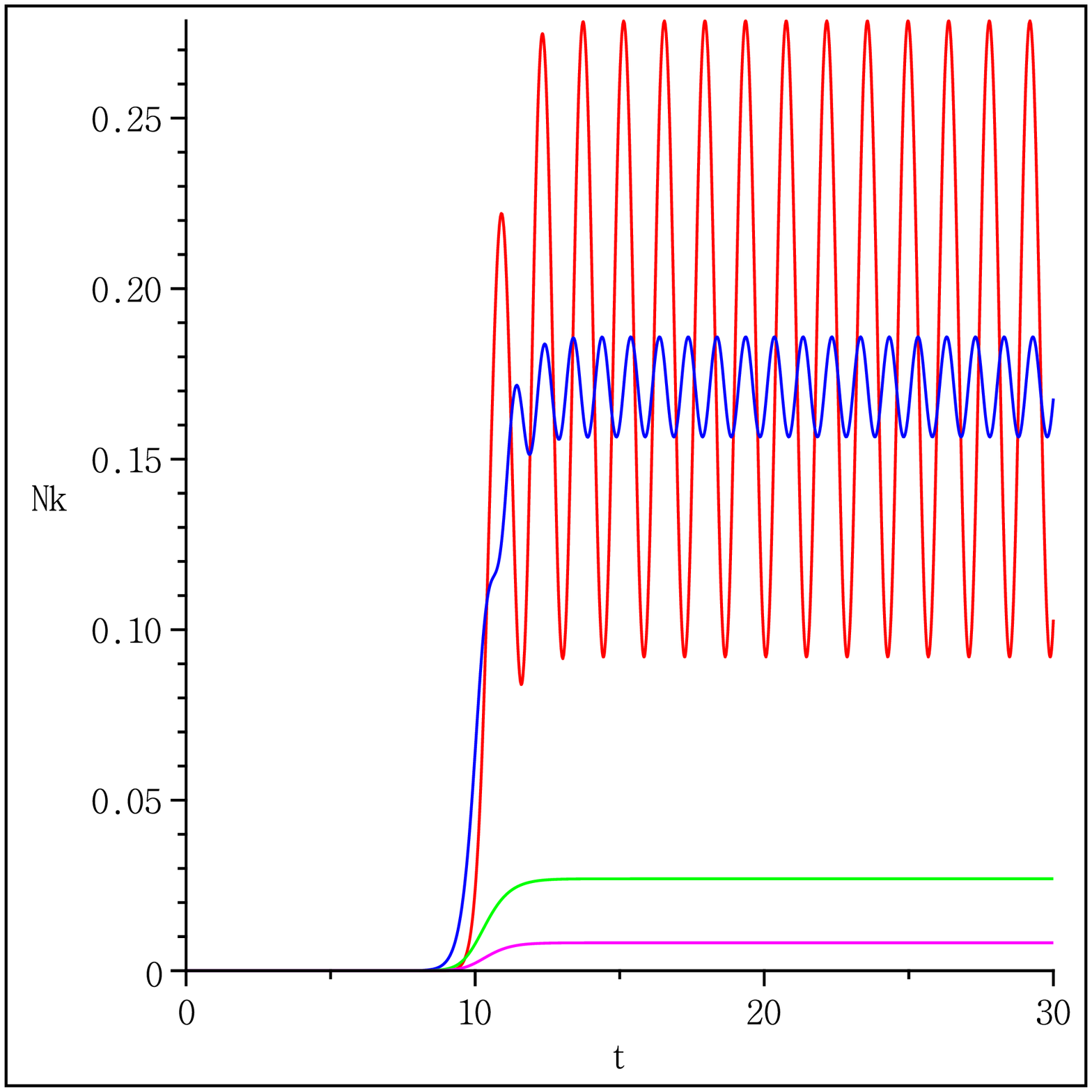}}\hfill
{\includegraphics[width=0.475\linewidth]{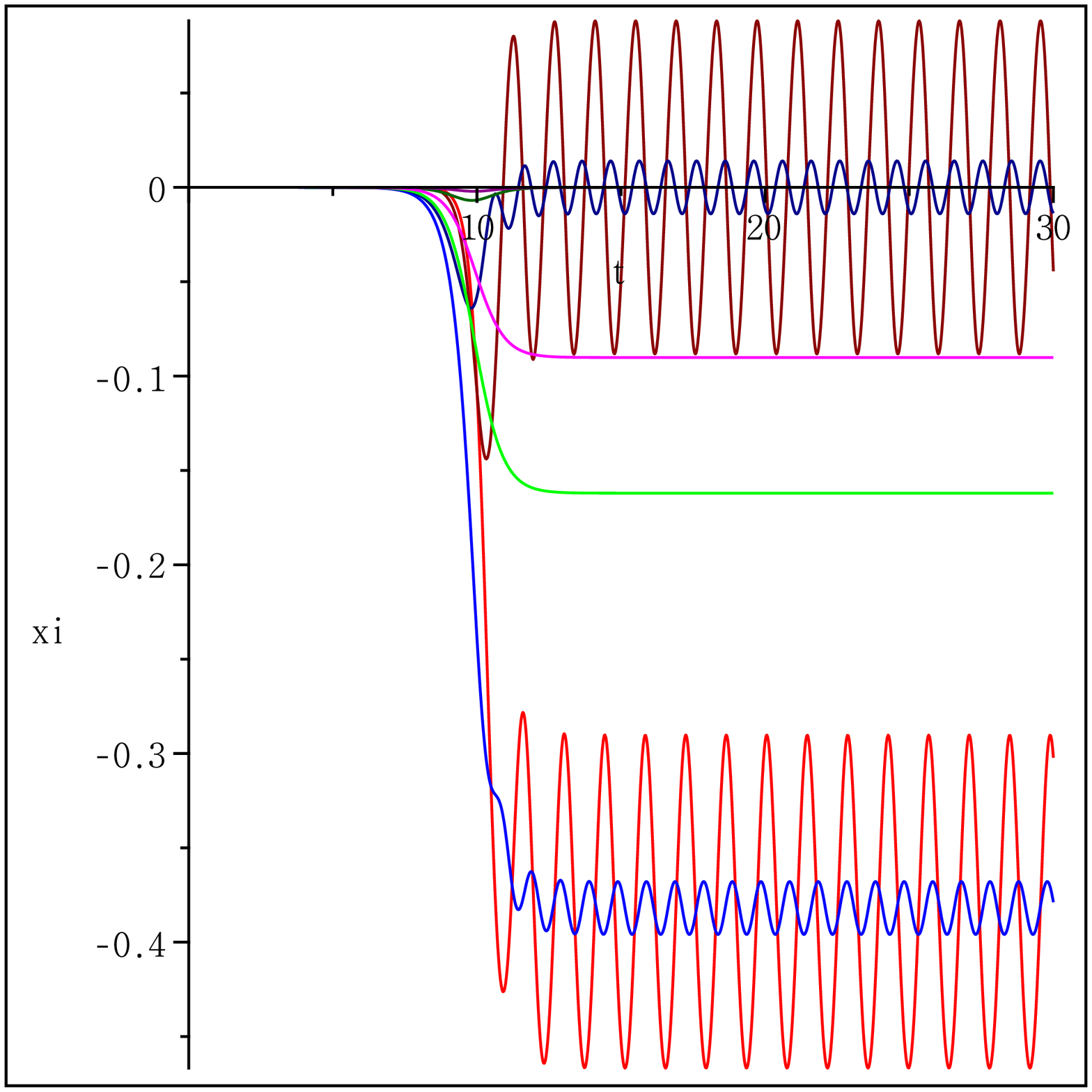}}
\caption{(color online). For the Sauter electric field, the number of pairs $N_k (t)$ [left panel] and the real and imaginary parts of
the function $\xi_k (t)$ [right panel] are plotted as a function of time
for $k_{\parallel} = 0$ (red),
$k_{\parallel} = 1$ (blue), $k_{\parallel} = 5$ (green), and $k_{\parallel} = 10$ (magenta), where
the real part is in the light color and the imaginary part is in the dark color. } \label{tanh-F2}
\end{figure}
\begin{figure}[t]
{\includegraphics[width=0.475\linewidth]{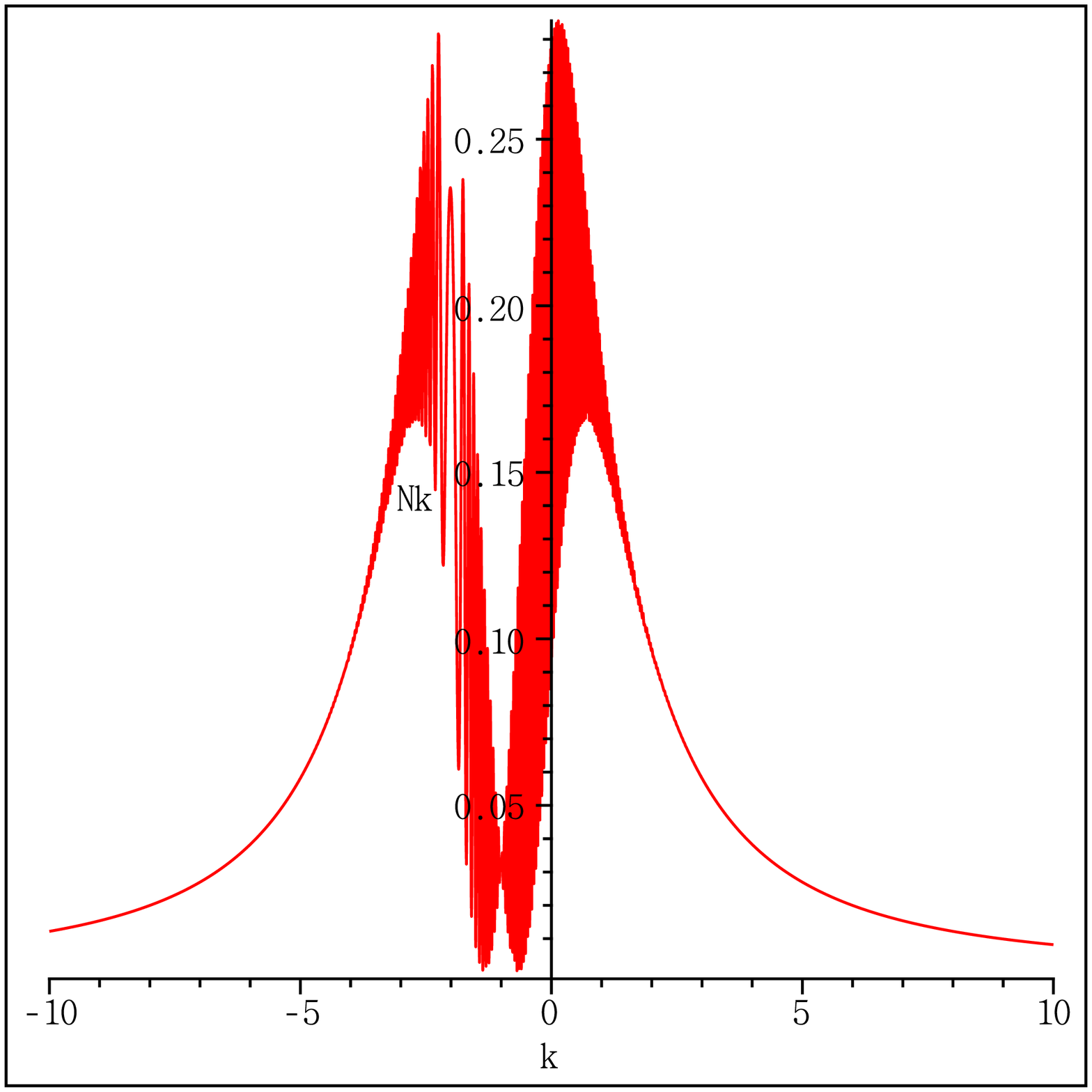}}\hfill
{\includegraphics[width=0.475\linewidth]{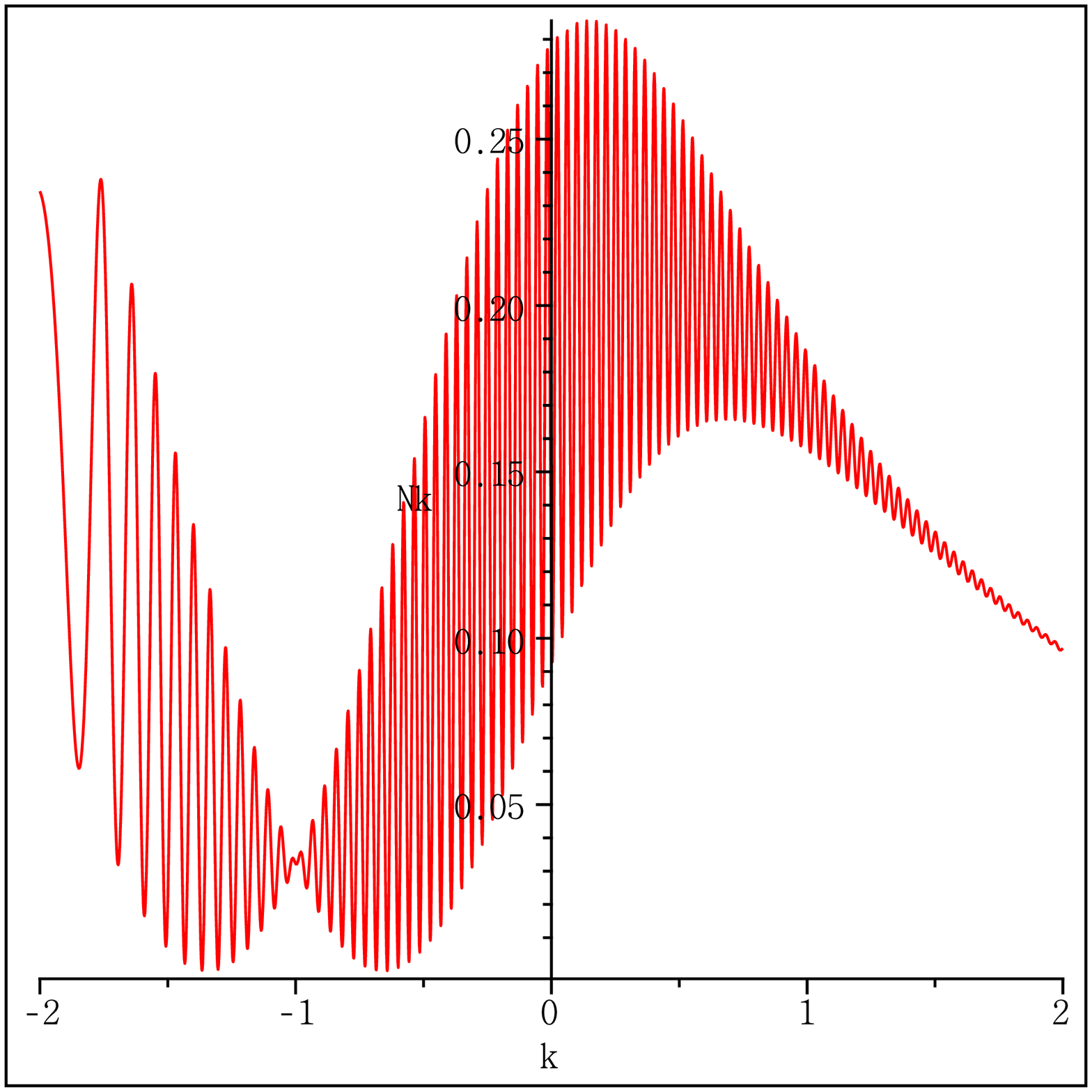}}
\caption{(color online). For the Sauter electric field, the longitudinal momentum spectrum of pairs at the time $t = 100$ is plotted
as a function of momentum in the range of $-10 \leq k_{\parallel} \leq 10$ [left panel] and
magnified in the range of $-2 \leq k_{\parallel} \leq 2$ [right panel].} \label{tanh-F3}
\end{figure}

At the critical strength $E_0 = 1$ for a Compton time scale pulse $\tau =1$,
the number of pairs is order of one and exhibits both the temporal behavior
and the substructure of the longitudinal momentum spectrum as shown in Fig.\,\ref{tanh-F1}.
The pair production for small momentum in the Compton unit begins to oscillate
around the time of interaction as shown in detail in Fig.\,\ref{tanh-F2},
in which the gauge potential at $t = 10$ is the half of the asymptotic value
in the remote future.
The pair production increases soon after the electric field is turned on, reaches the maximum within a few Compton times
and then oscillates with large amplitude for small $k_{\parallel}$ and with small amplitude for large $k_{\parallel}$.
The small amplitude oscillation for large $k_{\parallel}$ is not shown in Fig.\,\ref{tanh-F1} and in
the left panel of Fig.\,\ref{tanh-F2} due to a different order of magnitude.
However, the asymptotic solutions (\ref{asym sol2}) predict oscillations for all $k_{\parallel}$
since $A_{\parallel} (\infty) \neq 0$, which is  numerically confirmed.
The time-averaged number of pairs monotonically decreases as the momentum increases.

The suppression of pair production for large momentum is expected from the scattering picture in Sec. IV, according to which the particle has
a large kinetic energy compared to the potential well or barrier and thus has a small reflection coefficient, implying
small pair production. The oscillatory and temporal behavior of the squeezed vacuum (\ref{sq vac})
can understood from $\xi_k$ in the right panel of Fig.\,\ref{tanh-F2},
in which it oscillates with large amplitude for small momentum while it oscillates with small amplitude for large momentum,
according to the asymptotic solutions (\ref{asym sol2}).

The longitudinal momentum spectrum of pairs shows a substructure, which could not be seen by the number of
adiabatic pairs \cite{Nikishov70}.
Since the kinematic momentum $P_{\parallel} (t)$ for $k_{\parallel} + 1$ and $-(k_{\parallel} + 1)$ is antisymmetric in time,
the longitudinal momentum spectrum of number of pairs is symmetric around $k_{\parallel} = -1$ according to
Sec.\,\ref{scat pic}, which is numerically confirmed in Fig.\,\ref{tanh-F3}.
Pairs are minimally produced for $k_{\parallel} = -1$, for which $P_{\parallel} (-t) = - P_{\parallel} (t)$
and $\omega_k (- \infty) = \omega_k (\infty)$. The pair production is suppressed for large momentum
as expected.

\subsection{$E(t) = E_0 e^{- \frac{t^2}{\tau^2} }$} \label{Gaus}
\begin{figure}[ht]
\includegraphics[width=0.75\linewidth]{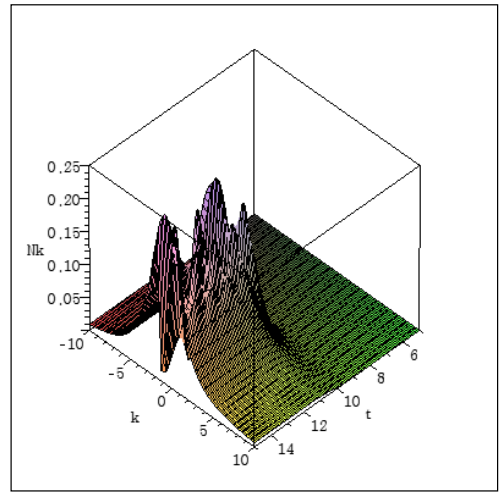}
\caption{(color online). The number of pairs in the Gaussian electric field with $E_0 = 1$ and $\tau =1$
is plotted as a function of time and longitudinal momentum.}
\label{erf-F1}
\end{figure}
\begin{figure}[t]
{\includegraphics[width=0.475\linewidth]{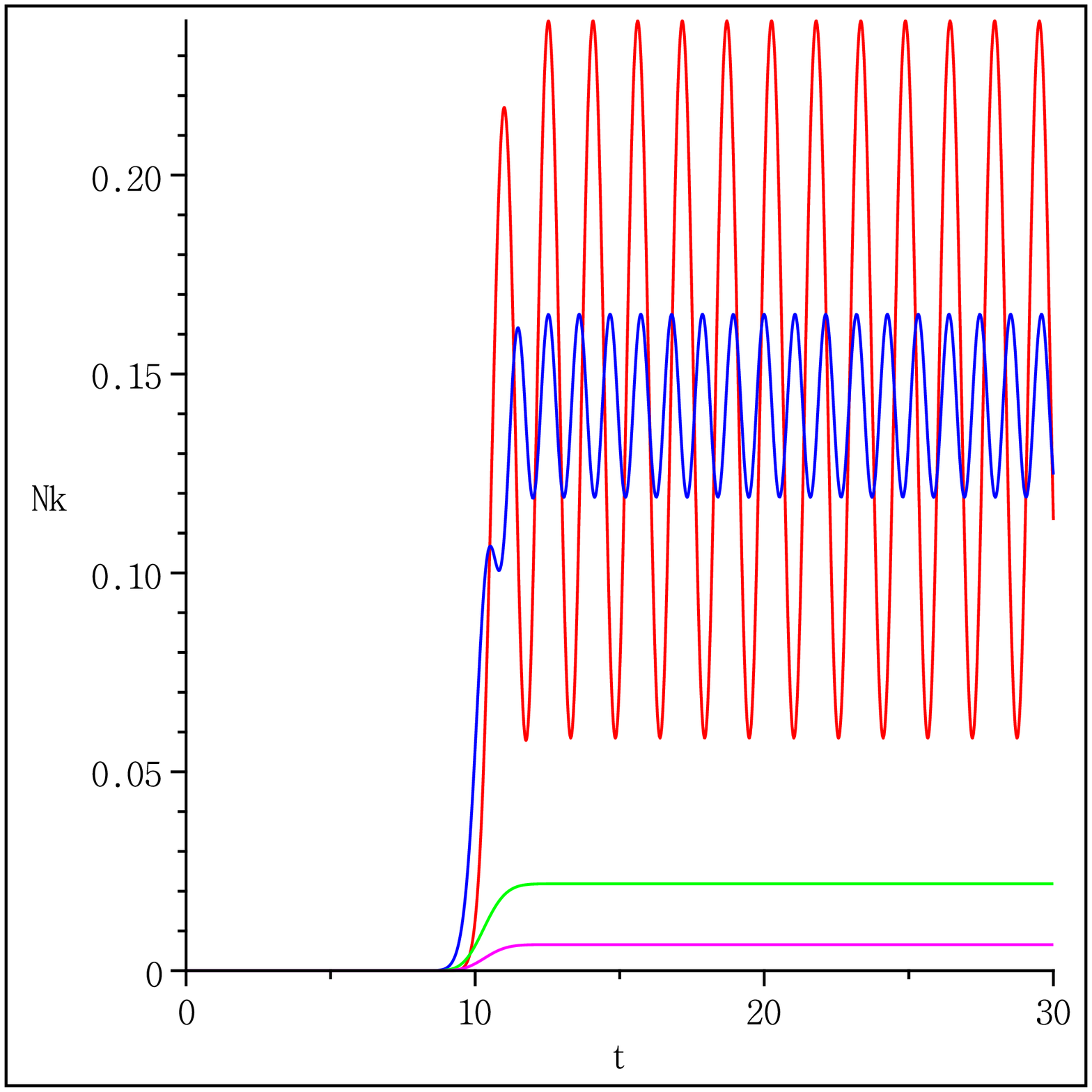}}\hfill
{\includegraphics[width=0.475\linewidth]{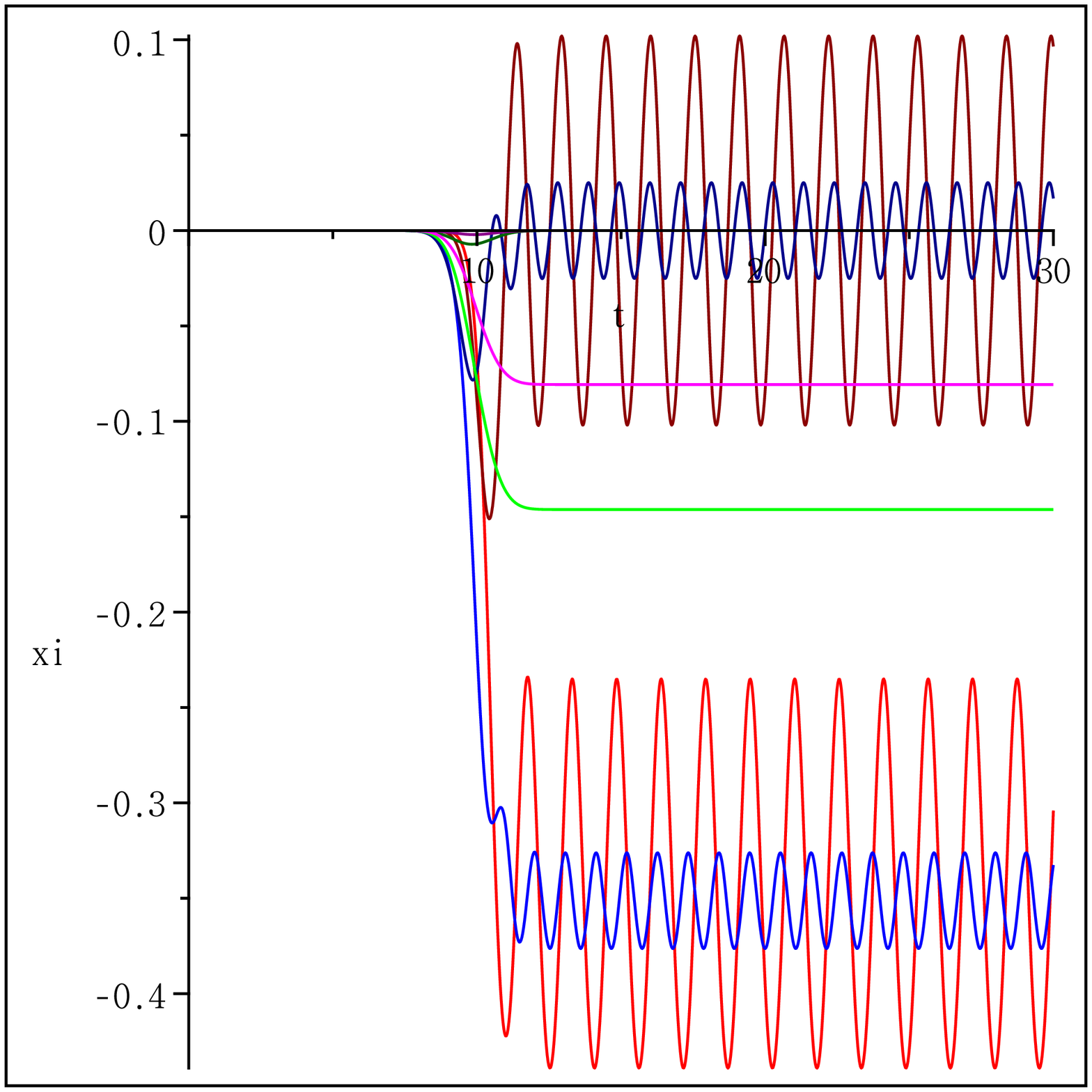}}
\caption{(color online). For the Gaussian electric field, the number of pairs $N_k (t)$ [left panel] and the real and imaginary parts of
the function $\xi_k (t)$ [right panel] are plotted as a function of time
for $k_{\parallel} = 0$ (red), $k_{\parallel} = 1$ (blue), $k_{\parallel} = 5$ (green), and $k_{\parallel} = 10$ (magenta),
where the real part is in the light color and the imaginary part is in the dark color.} \label{erf-F2}
\end{figure}
\begin{figure}[t]
{\includegraphics[width=0.475\linewidth]{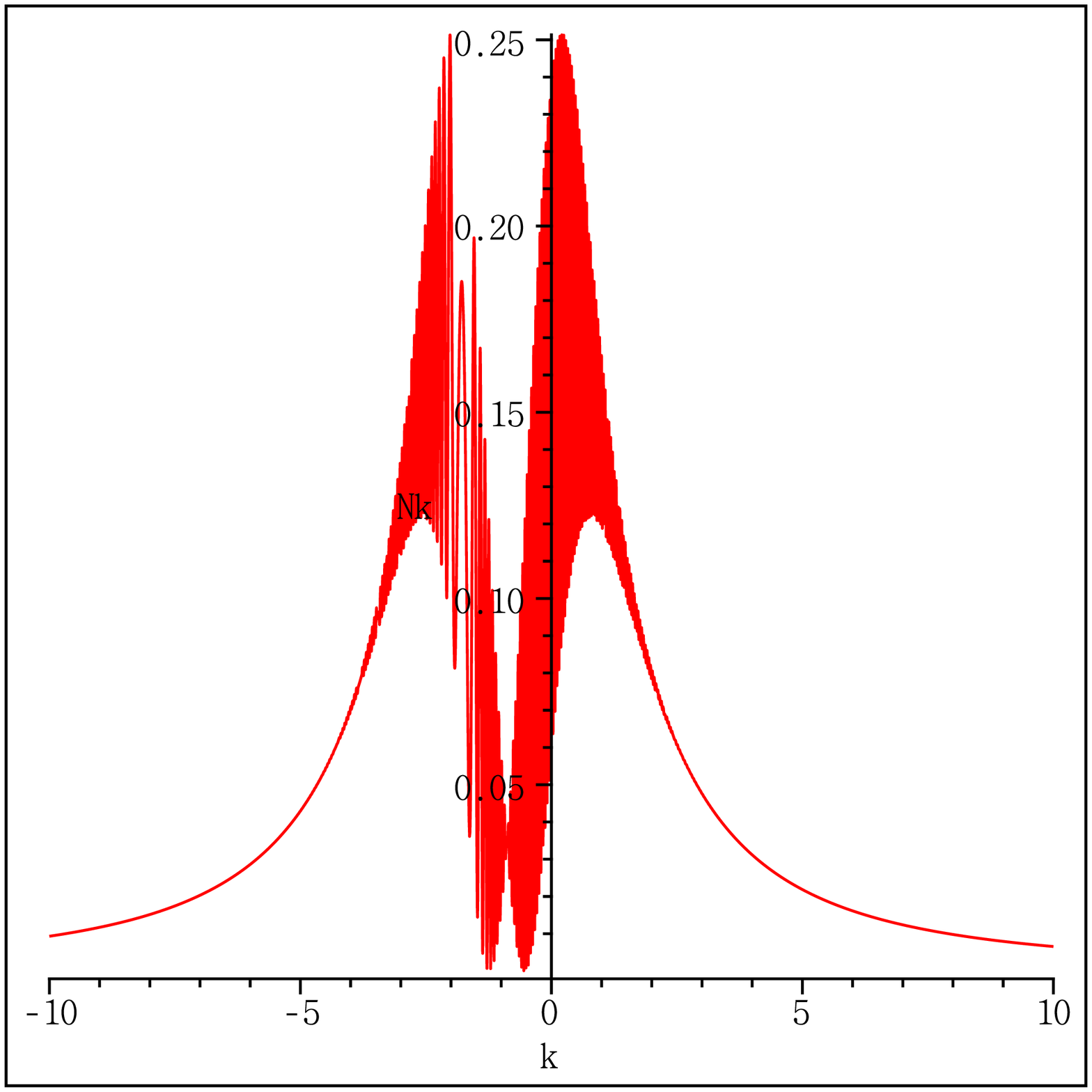}}\hfill
{\includegraphics[width=0.475\linewidth]{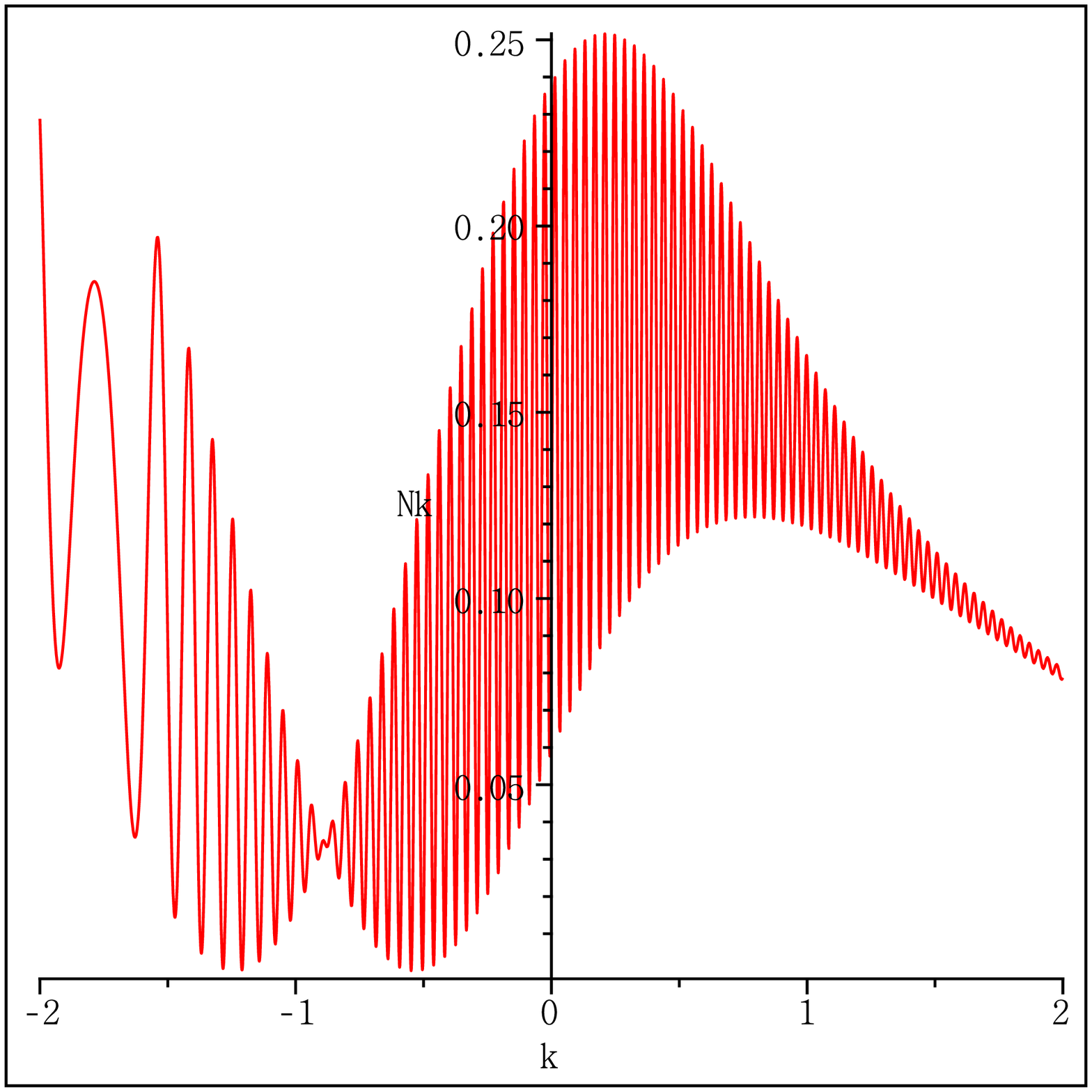}}
\caption{(color online). For the Gaussian electric field, the longitudinal momentum spectrum of pairs at $t = 100$ is plotted
in the range of $-10 \leq k_{\parallel} \leq 10$ [left panel] and magnified
in the range of $-2 \leq k_{\parallel} \leq 2$ [right panel].} \label{erf-F3}
\end{figure}

The gauge potential given by the error-function
\begin{eqnarray}
A_{\parallel} (t) &=& - E_0 \left ( \frac{\sqrt{\pi}\tau}{2} + \int_{0}^{t -10 \tau} e^{- \frac{t'^2}{\tau^2} }dt' \right )
\nonumber\\&=& - \frac{\sqrt{\pi}E_0 \tau}{2}
\left \{ 1 + {\rm erf} \left (\frac{t- 10\tau}{\tau}\right ) \right \},
\end{eqnarray}
leads to the Gaussian electric field
\begin{eqnarray}
E (t) = E_0 e^{- \frac{(t-10 \tau)^2}{\tau^2} }.
\end{eqnarray}
The center of the Gaussian field is shifted for the numerical purpose.
The Gaussian electric field decays more rapidly and thus produces
relatively smaller pairs than the Sauter electric field in Sec.\,\ref{Saut}.
For the numerical work we set $E_0 = 1$ and $\tau =1$ as for the Sauter electric field.

The number of pairs as function of time and longitudinal momentum in Fig.\,\ref{erf-F1}
exhibits a similar structure as that of the Sauter field in Fig.\,\ref{tanh-F1}.
It has a structure for small momentum but it is suppressed for large momentum.
The similarity of the temporal behavior can be seen by comparing the left panel of Fig.\,\ref{erf-F2} with Fig.\,\ref{tanh-F2}.
The longitudinal momentum spectrum of pairs in Fig.\,\ref{erf-F3} also has a similar pattern as Fig.\,\ref{tanh-F3}.
As the error-function is an odd function ${\rm erf} (-t) = - {\rm erf} (t)$, it is
symmetric around $k_{\parallel} = - \sqrt{\pi}/2$ according to Sec.\,\ref{scat pic},
which is numerically confirmed in Fig.\,\ref{erf-F3}. Note that
$k_{\parallel} = - \sqrt{\pi}/2$ is the channel for minimal pair production for small momentum.
The suppression of pair production  for large momentum is also shown in Fig.\,\ref{erf-F3}.

\subsection{$A_{\parallel} (t) = \frac{E_0}{\omega}  e^{- \frac{t^2}{\tau^2}} \cos(\omega t)$} \label{Gaus-cos}
\begin{figure}[ht]
\includegraphics[width=0.475\linewidth]{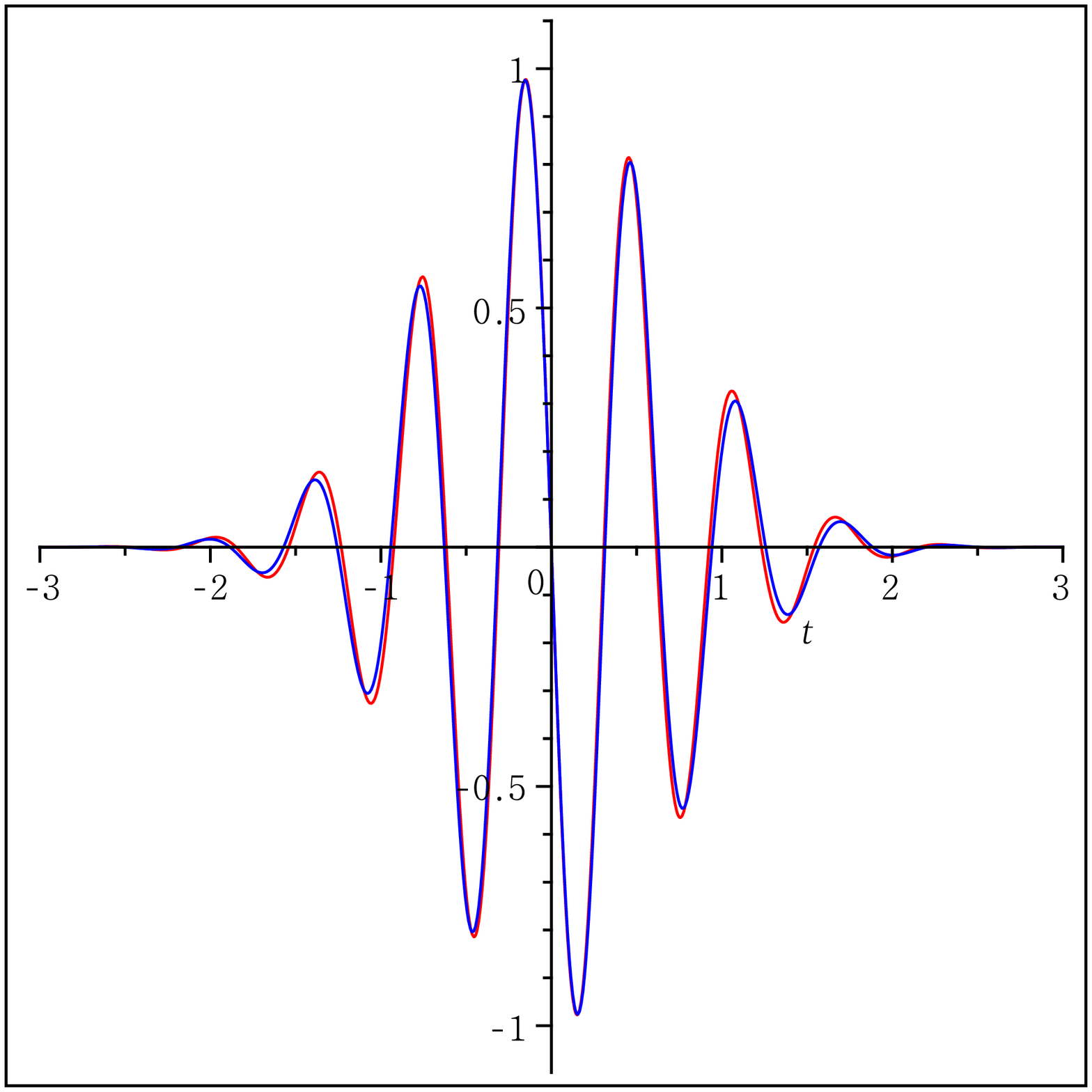}\hfill
\includegraphics[width=0.475\linewidth]{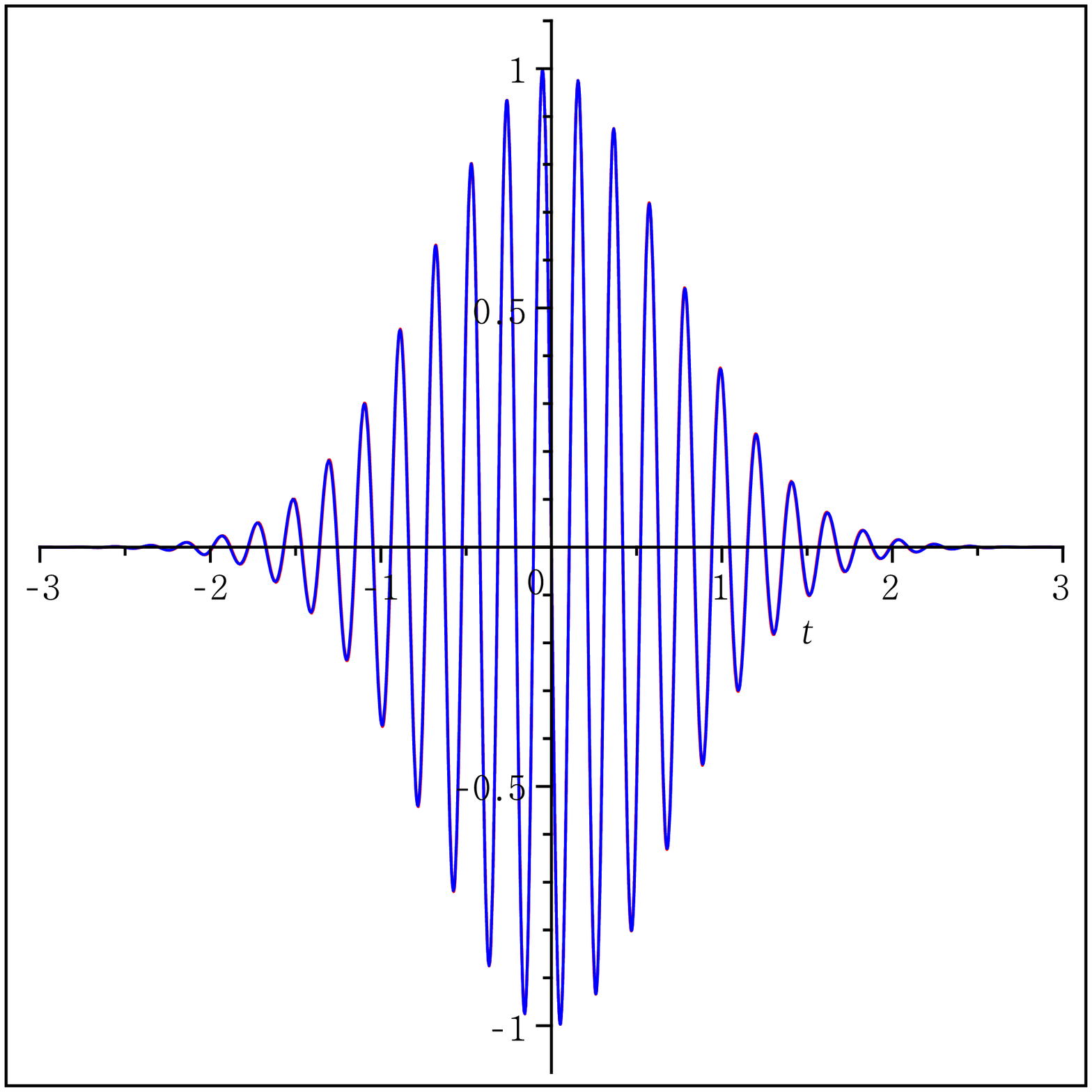}
\caption{(color online). The profile of electric field (\ref{Gaus-cos-E})
is plotted in red color for $E_0 = 1$, $\tau =1$ and $\omega = 10$ [left panel]
and $\omega = 30$ [right panel]. The curve in blue color in each panel is $e^{- (t-10)^2} \sin(\omega t)$ for the same value of
$\omega$.}
\label{Gcos-F1}
\end{figure}
\begin{figure}[ht]
\includegraphics[width=0.75\linewidth]{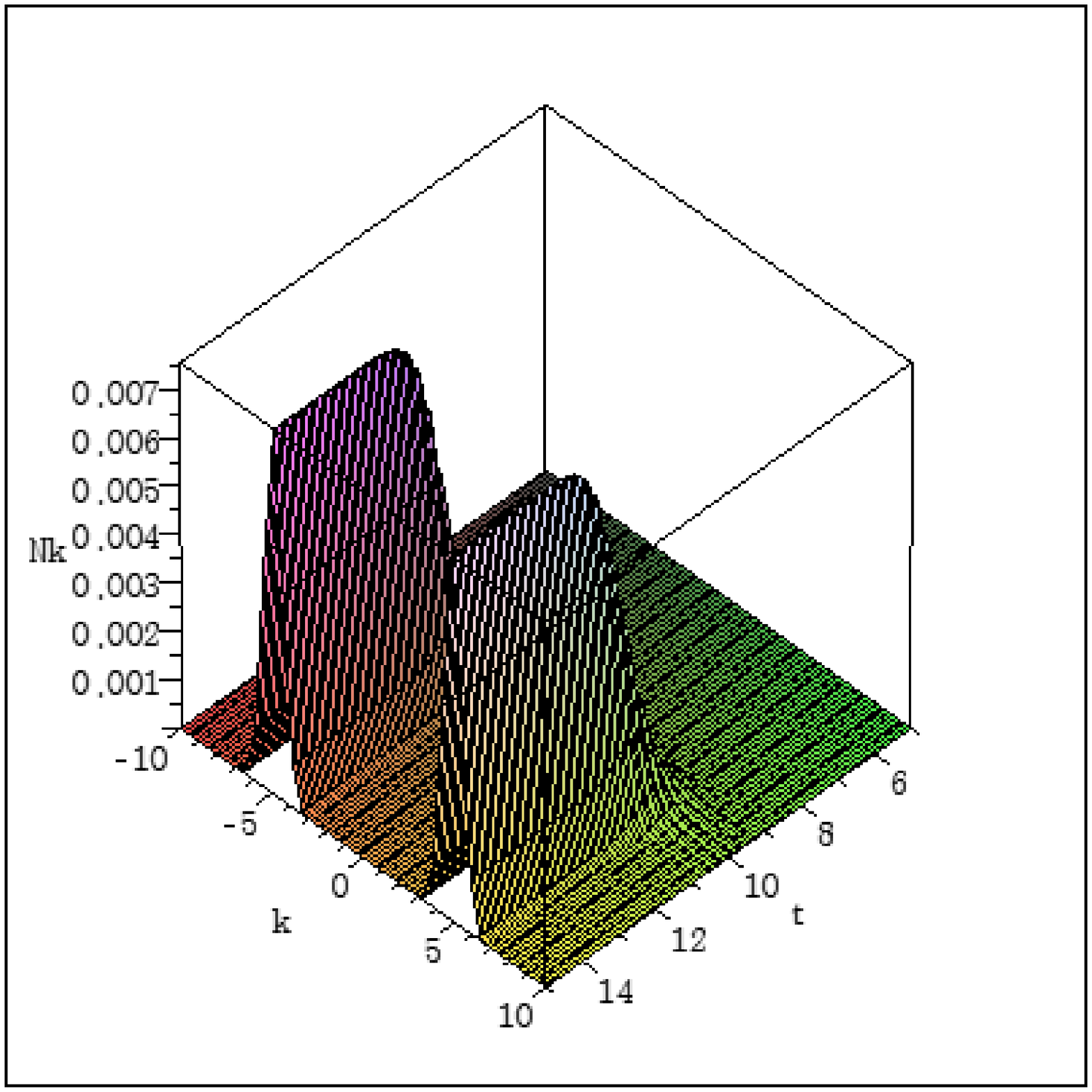}
\caption{(color online). The number of pairs in the oscillating Gaussian electric field with $E_0 = 1$ and $\tau =1$
is plotted as a function of time and longitudinal momentum.}
\label{Gcos-F2}
\end{figure}
\begin{figure}[t]
{\includegraphics[width=0.475\linewidth]{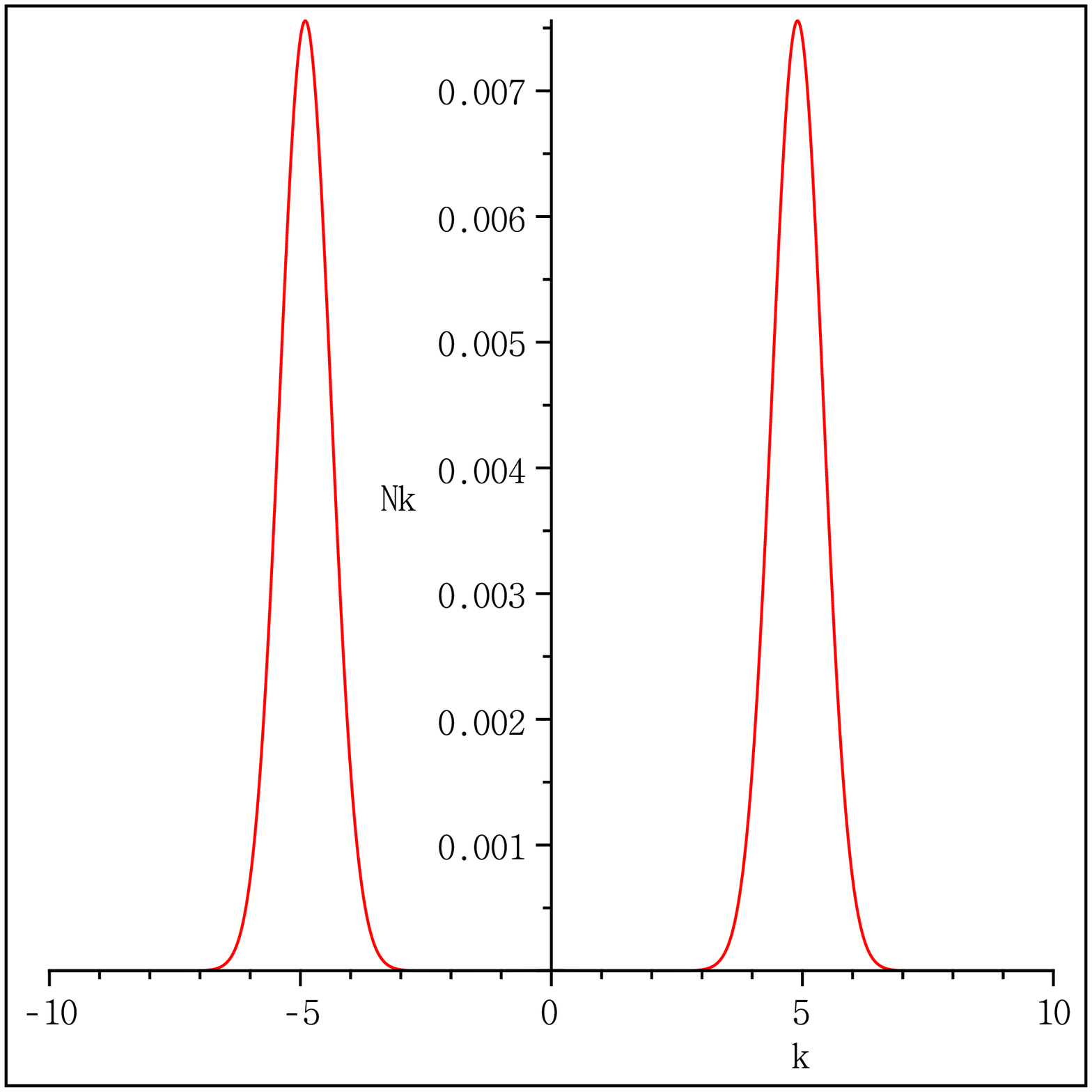}}\hfill
{\includegraphics[width=0.475\linewidth]{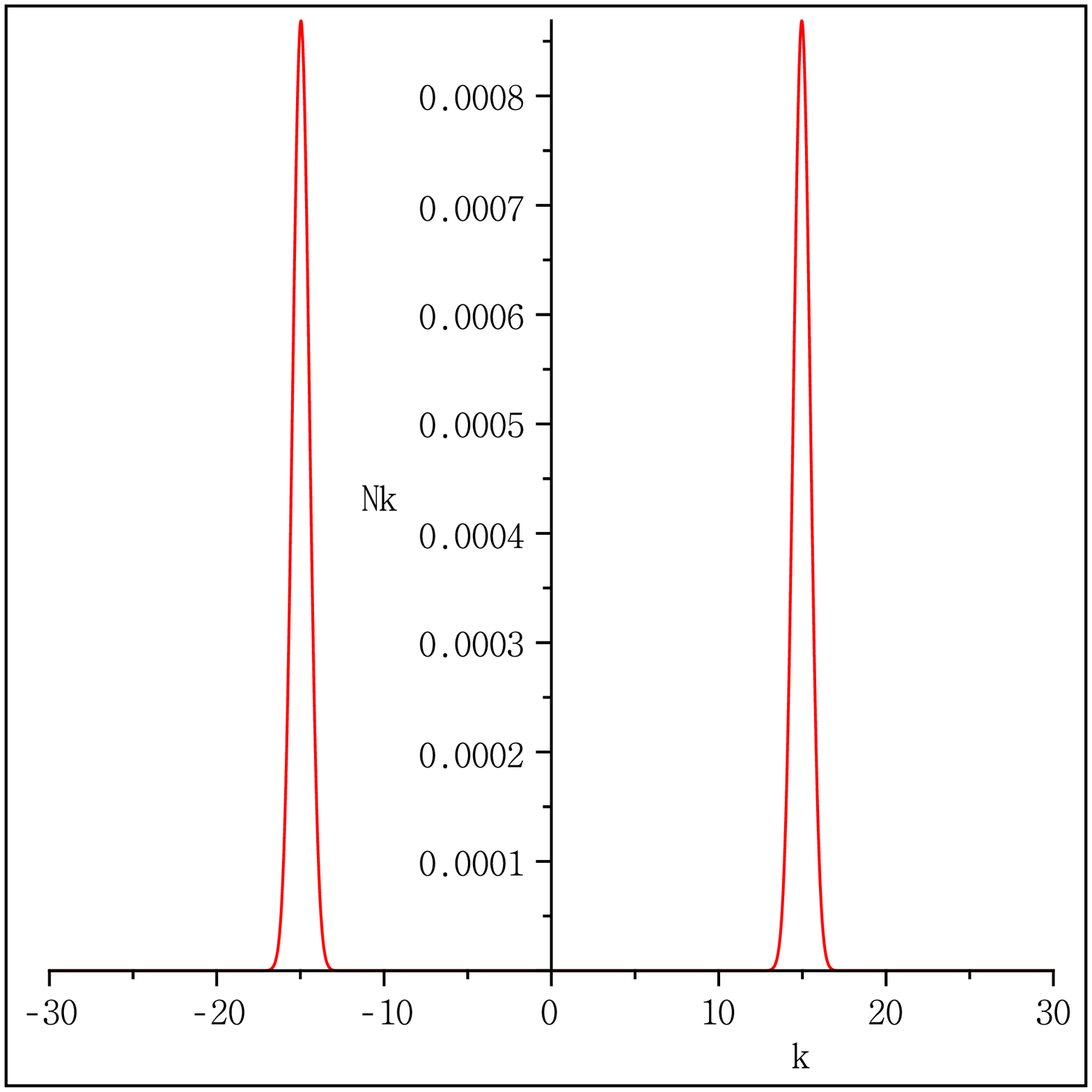}}
\caption{(color online). For the oscillating Gaussian electric field,
the longitudinal momentum spectrum of pairs is plotted at $t = 100$ for $\omega = 10$ [left panel]
and $\omega = 30$ [right panel].} \label{Gcos-F3}
\end{figure}

Finally, we consider a gauge potential
\begin{eqnarray}
A_{\parallel} (t) = \frac{E_0}{\omega}  e^{- \frac{(t-10 \tau)^2}{\tau^2}} \cos(\omega t),
\end{eqnarray}
which leads to the oscillating Gaussian electric field
\begin{eqnarray}
E (t) &=& E_0  e^{- \frac{(t-10 \tau)^2}{\tau^2}} \sin(\omega t)
\nonumber\\
&& + \frac{2 E_0}{\omega \tau} \Bigl(\frac{t- 10 \tau}{\tau} \Bigr) e^{- \frac{(t-10 \tau)^2}{\tau^2}} \cos(\omega t).
\label{Gaus-cos-E}
\end{eqnarray}
As shown in Fig.\,\ref{Gcos-F1} the first term is dominant in the region $|t - 10 \tau| \leq 2 / (\omega \tau^2)$
and oscillates with an Gaussian envelope $e^{- (t -10 \tau)^2/\tau^2}$
and beyond that region the second term has
a linearly growing factor but both terms are exponentially suppressed.

For the numerical work we set $E_0 = 1$, $\tau = 1$ and $\omega = 10$ or $\omega = 30$. It is surprising that
the number of pairs in Fig.\,\ref{Gcos-F2} is bunched around $k_{\parallel} = -5$ and $k_{\parallel} = 5$.
The longitudinal momentum spectrum in Fig.\,\ref{Gcos-F3} shows bunching with momentum separation of $10$ and $30$ for
$\omega = 10$ and $\omega = 30$, respectively. And the
rapidly oscillating factor suppresses pair production, though the peak intensity in Fig.\,\ref{Gcos-F1}
is almost comparable to the Sauter electric field in
Sec.\,\ref{Saut} and the Gaussian electric field in Sec.\,\ref{Gaus}.
The numerical study also shows a similar pattern in pair production and
bunching of pairs in the gauge potential
\begin{eqnarray}
A_{\parallel} (t) = \frac{E_0}{\omega}  e^{- \frac{(t-10 \tau)^2}{\tau^2}} \sin(\omega t).
\end{eqnarray}

\section{Pair Production in Di-Polarity Electric Fields} \label{di-pol}

In this section we study how the polarity of the electric field influences pair production, in particular, when
the electric field changes the polarity during the interaction. It is interesting to understand
how the produced pairs behave when another electric pulse of opposite polarity acts. Do they annihilate each
other partially or completely? Or are there more pairs produced by the second pulse?
In order to answer some of these questions we consider three model fields.

\subsection{$E(t) = \frac{E_0 }{\cosh^2 (\frac{t-t_1}{\tau})} - \frac{E_0 }{\cosh^2 (\frac{t-t_2}{\tau})}$} \label{di-Saut}

As the first model we consider the di-polarity Sauter electric field,
two Sauter electric fields acting with a time lag and in opposite directions.
The model gauge field is given by
\begin{eqnarray}
A_{\parallel} = - E_0 \tau \Bigl(\tanh (\frac{t-10 \tau}{\tau}) -  \tanh (\frac{t-20 \tau}{\tau})\Bigr).
\end{eqnarray}
Though we cannot exactly solve QED problem in this gauge potential, we may understand the characteristic
feature of pair production by each of Sauter electric fields when they are separated by a sufficient
time gap as in quantum mechanics.
\begin{figure}[ht]
\includegraphics[width=.75\linewidth]{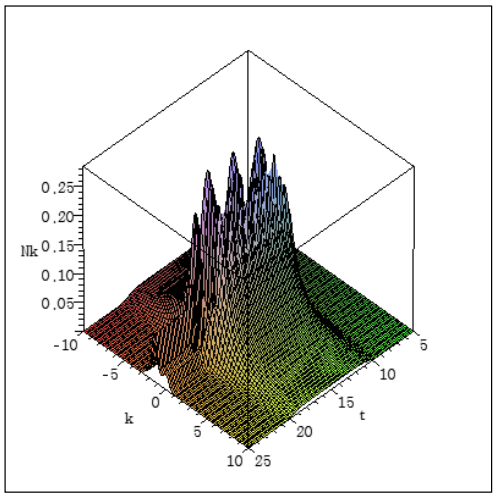}
\caption{(color online). The number of pairs in the di-polarity Sauter electric field with $E_0 = 1$ and $\tau =1$
is plotted as a function of time and longitudinal momentum.}
\label{disaut-F1}
\end{figure}
\begin{figure}[t]
{\includegraphics[width=0.475\linewidth]{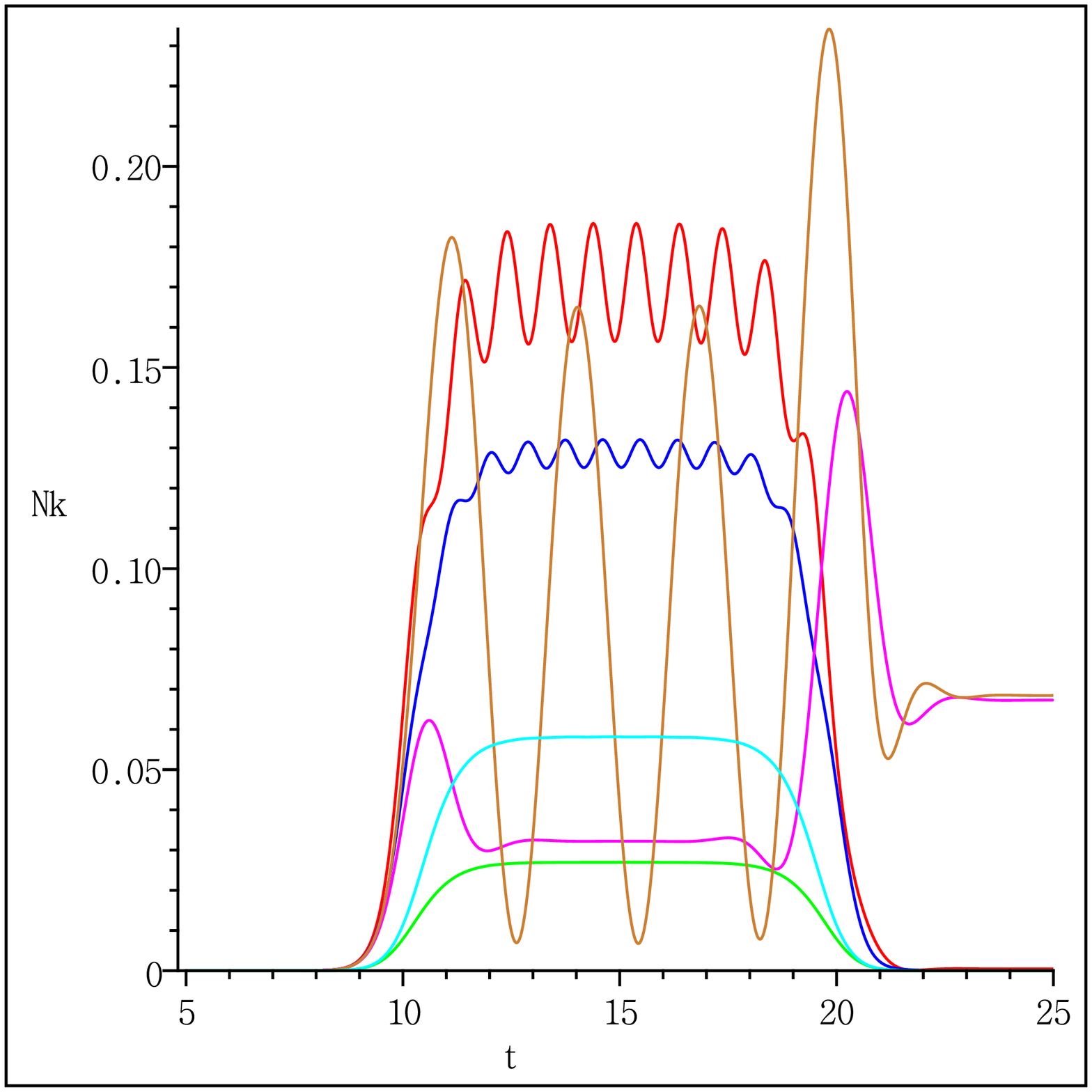}}\hfill
{\includegraphics[width=0.475\linewidth]{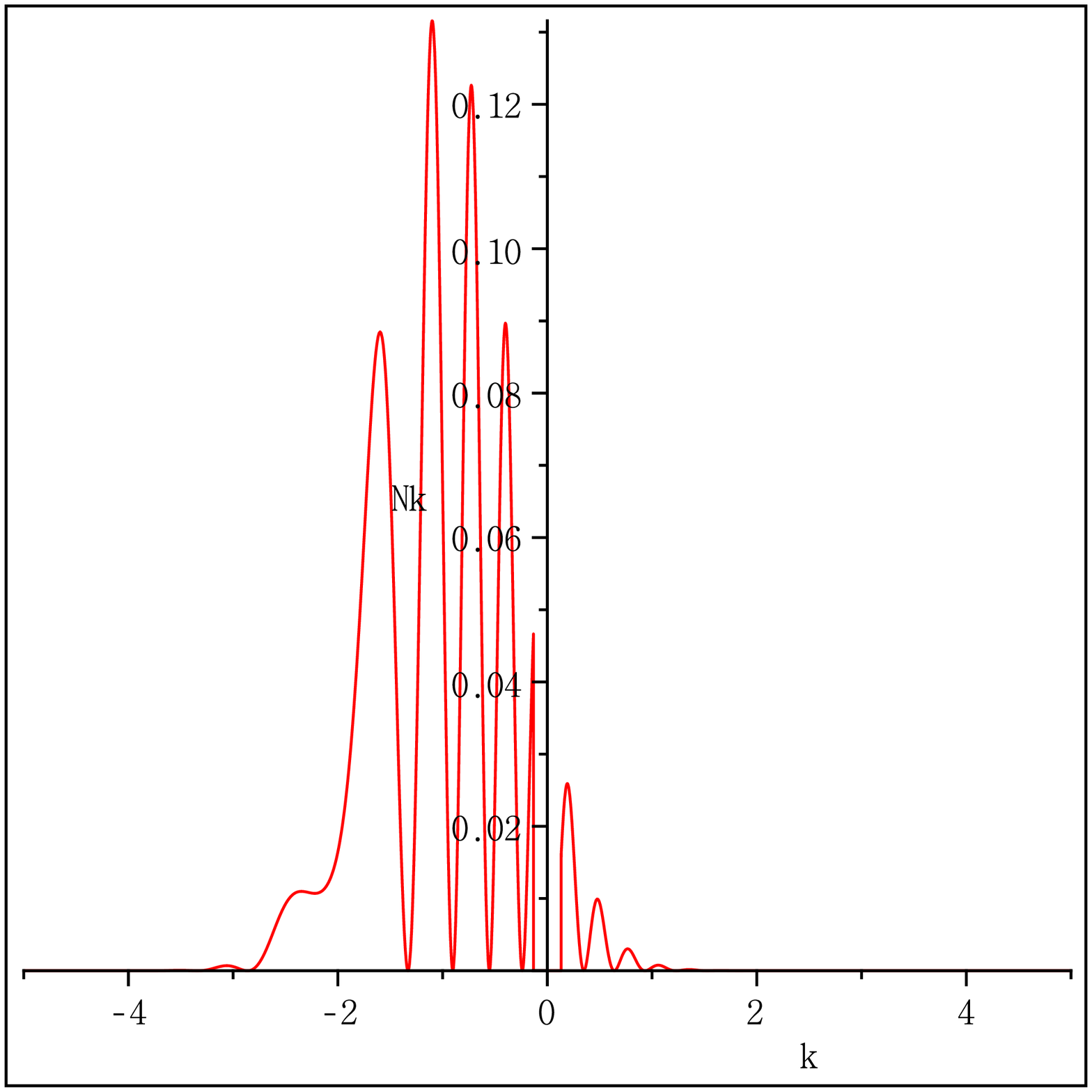}}
\caption{(color online). The number of pairs $N_k (t)$ in the di-polarity Sauter electric field
is plotted as a function of time for $k_{\parallel} = 1$ (red), $k_{\parallel} = 1.5$ (blue), $k_{\parallel} = 5$ (green), $k_{\parallel} = -1$ (magenta), $k_{\parallel} = -1.5$ (gold), and $k_{\parallel} = -5$ (skyblue)
[left panel] and the longitudinal momentum spectrum is plotted at $t = 100$ [right panel].} \label{disaut-F2}
\end{figure}
\begin{figure}[t]
{\includegraphics[width=0.475\linewidth]{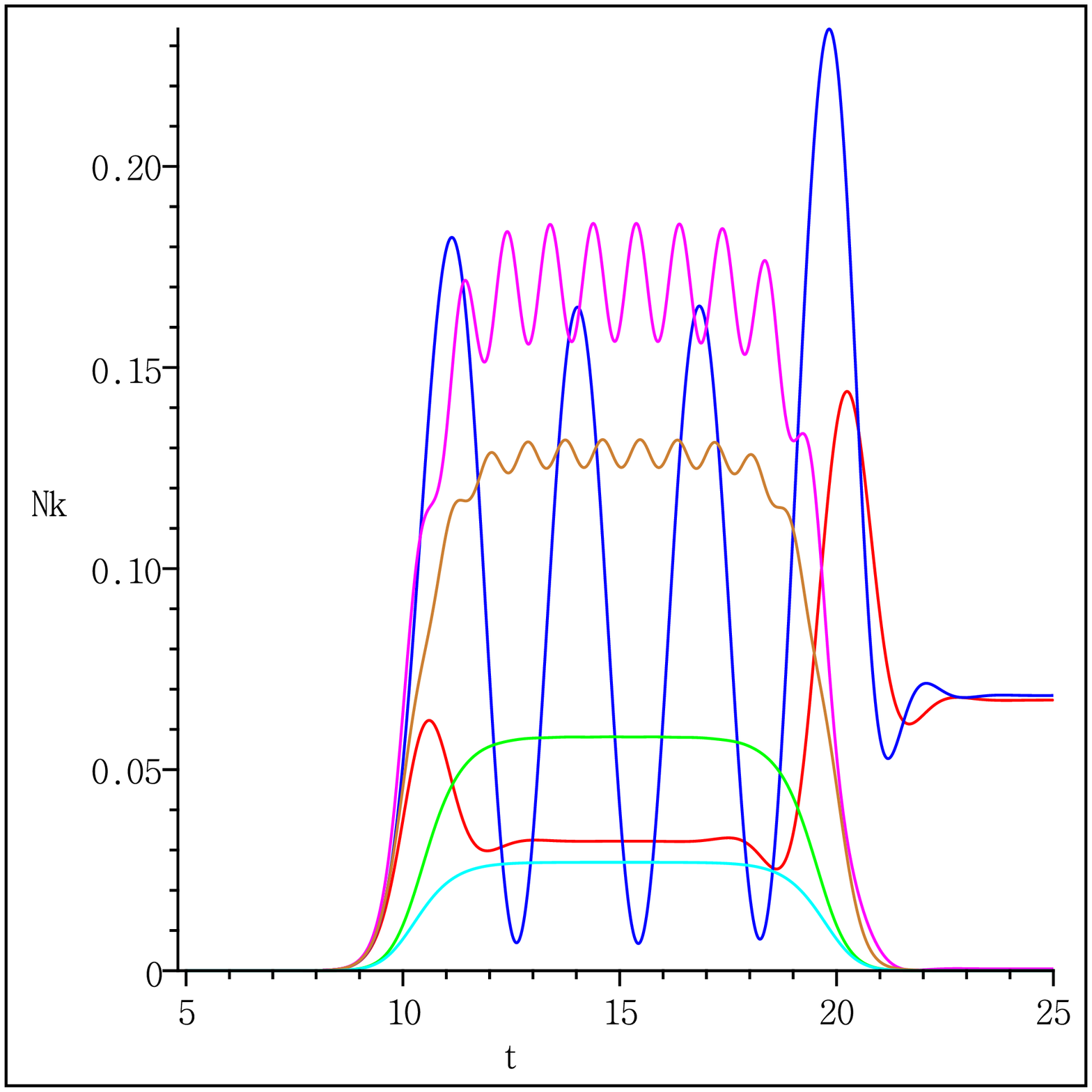}}\hfill
{\includegraphics[width=0.475\linewidth]{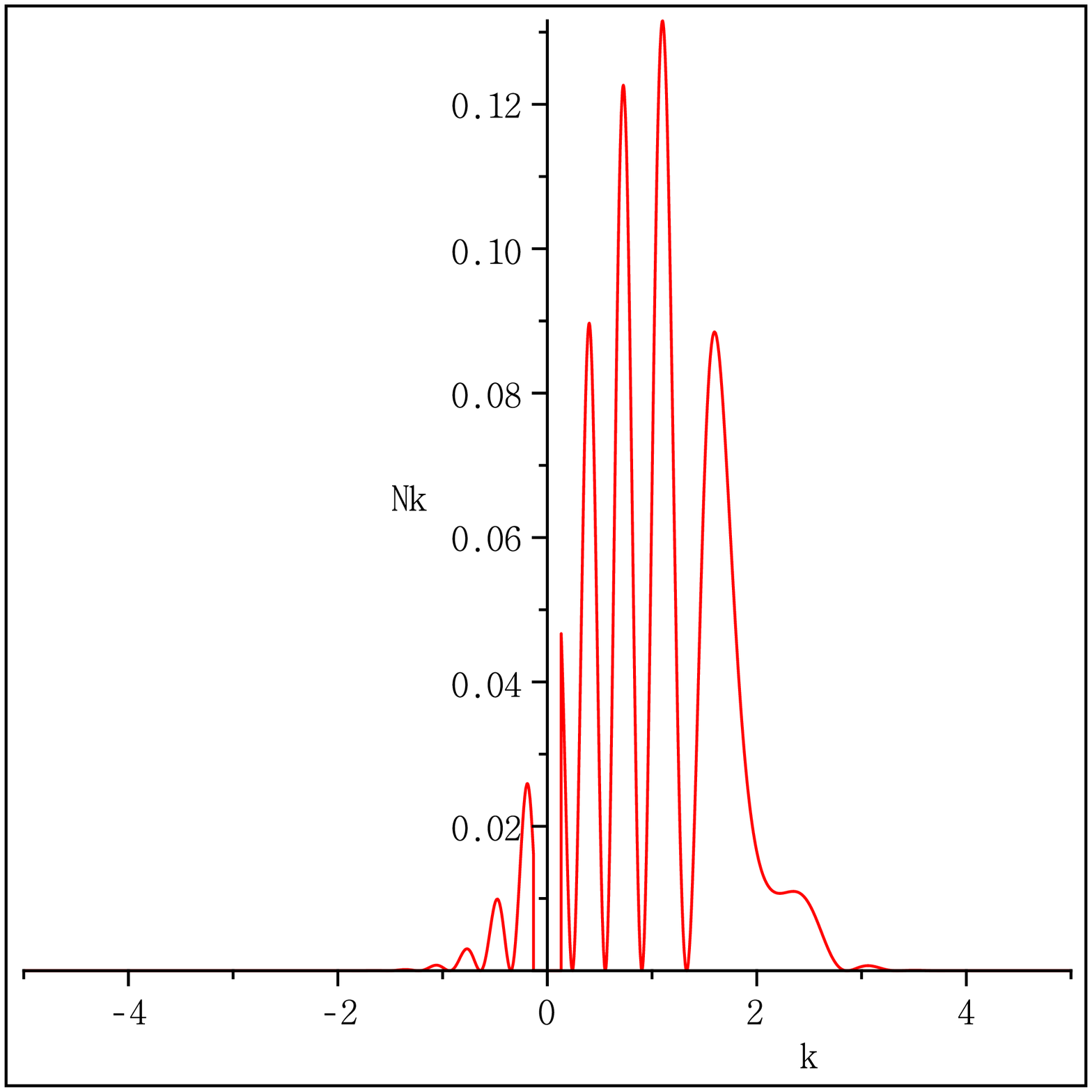}}
\caption{(color online). The number of pairs $N_k (t)$ in $E(t) = - \frac{1}{\cosh^2 (t-10)} + \frac{1}{\cosh^2 (t -20)}$
is plotted as a function of time for $k_{\parallel} = 1$ (red), $k_{\parallel} = 1.5$ (blue), $k_{\parallel} = 5$ (green), $k_{\parallel} = -1$ (magenta), $k_{\parallel} = -1.5$ (gold), and $k_{\parallel} = -5$ (skyblue)
[left panel] and the longitudinal momentum spectrum is plotted at $t = 100$ [right panel].} \label{disaut-F3}
\end{figure}

For the numerical work we set $E_0 = 1$ and $\tau =1$ so that two Sauter electric pulses are effectively
separated. At the onset of the interaction the peak of number of pairs in Fig.\,\ref{disaut-F1} is almost the same as the single
Sauter electric field in Fig.\,\ref{tanh-F1}, though it exhibits more structure. However, the main
difference appears after the completion of the interaction. The di-polarity Sauter electric field returns
to the Minkowski vacuum while the Sauter electric field has a constant residual gauge and $\omega_k (- \infty) \neq \omega_k (\infty)
$. The asymptotic solutions (\ref{asym sol1}) predict that the number of pairs for the di-polarity Sauter field
approaches to a constant, as shown in Figs.\,\ref{disaut-F1} and \ref{disaut-F2},
while Eq.\,(\ref{asym num2}) predicts that the number of pairs for the Sauter field
oscillates with a constant amplitude around the time-averaged value, as shown in Figs.\,\ref{tanh-F1} and
\ref{tanh-F2}.

The gauge potential in between two peaks of the di-polarity Sauter field has approximately
the shape of square potential barrier, in which pair production oscillates according to
the asymptotic solution (\ref{asym num2}). No pair production for $k_{\parallel} = 0$ is
the numerical coincidence that $m=1$ in
the Compton unit is equivalent to five wavelengths of a particle in ten Compton length width of the square
barrier in the scattering picture \ref{scat pic} and has the zero reflection coefficient due to resonance.

The effect of the polarity of the electric field can be seen in the right panel of Fig.\,\ref{disaut-F2} and
Fig.\,\ref{disaut-F3}, in which the longitudinal momentum spectrum shows the mirror symmetry when the polarity
of the electric field changes. There are residual pairs with negative momentum (Fig.\,\ref{disaut-F2})
when the first Sauter field acts in the positive direction and then the second one in the negative direction,
while more pairs with positive momentum survive (Fig.\,\ref{disaut-F3})
when the first Sauter field acts in the negative direction and then the second one in the positive direction.
The number of pairs in time also shows the polarity effect as shown in the left panel of Fig.\,\ref{disaut-F2} and
Fig.\,\ref{disaut-F3}.

\subsection{$A_{\parallel} (t) = \frac{E_0 \tau}{1 + \frac{t^2}{\tau^2}}$} \label{inv pot}
\begin{figure}[ht]
\includegraphics[width=0.75\linewidth]{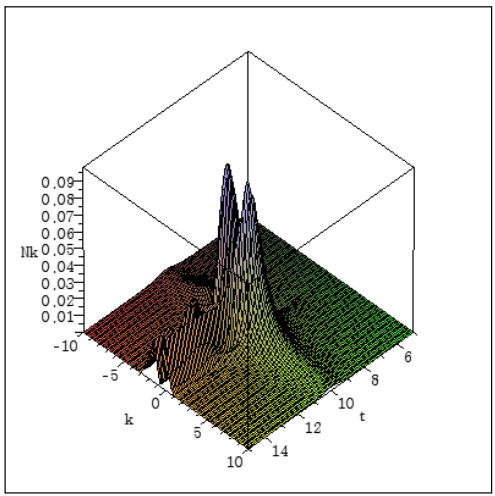}
\caption{(color online). The number of pairs in the inverse square potential
is plotted as a function of time and longitudinal momentum.}
\label{inv-F1}
\end{figure}
\begin{figure}[t]
{\includegraphics[width=0.475\linewidth]{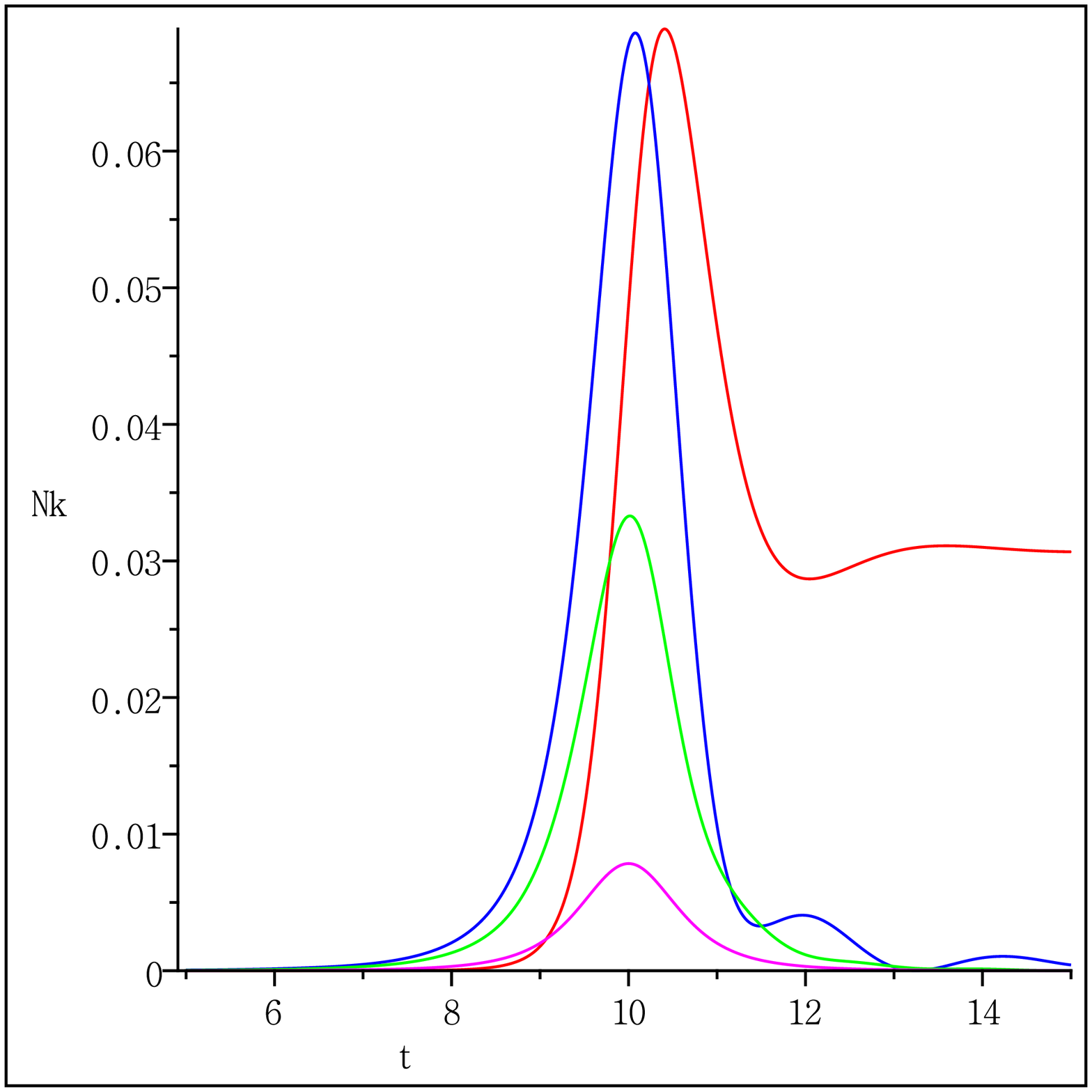}}\hfill
{\includegraphics[width=0.475\linewidth]{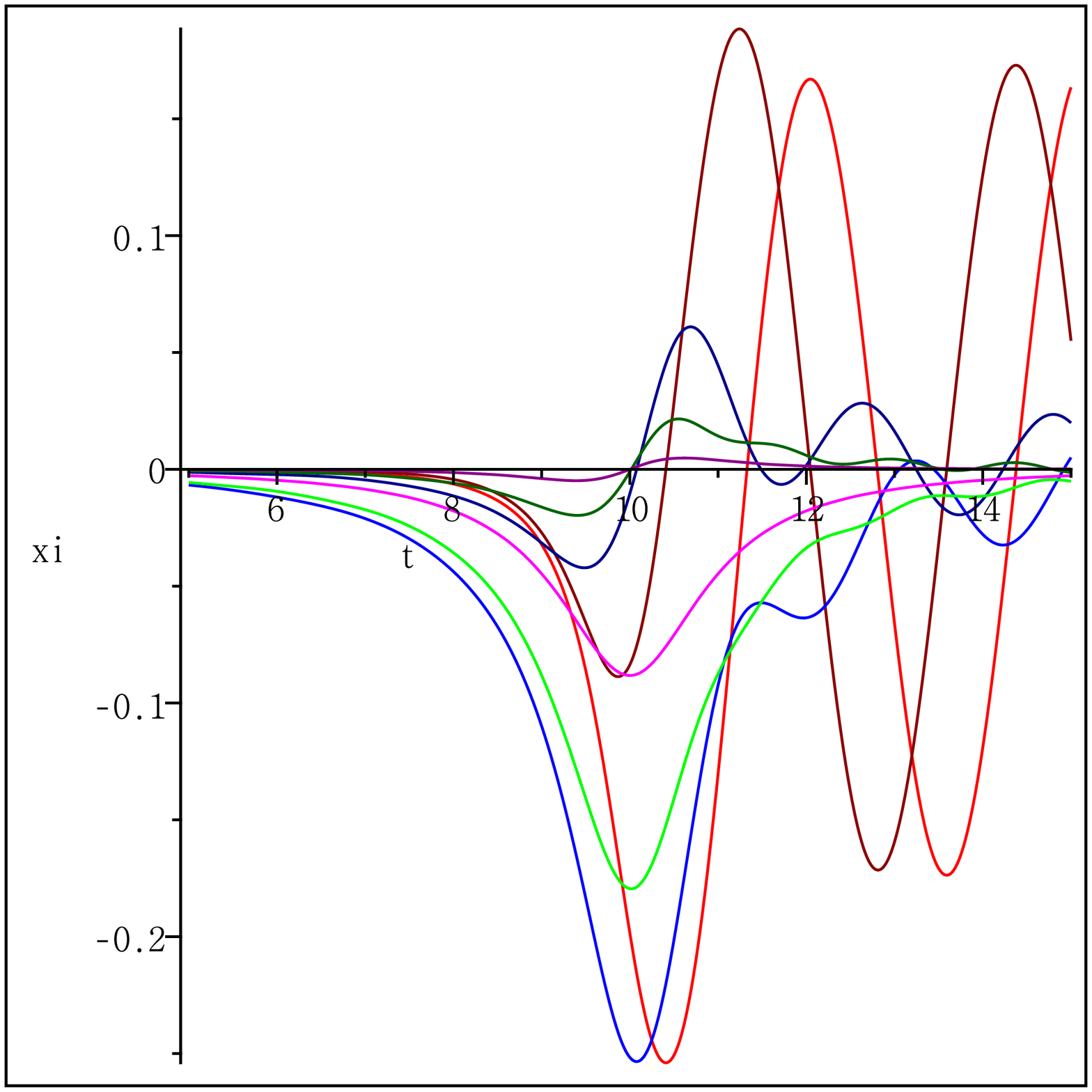}}
\caption{(color online). For the inverse square potential, the number of pairs $N_k (t)$ [left panel] and the real and imaginary part of
the function $\xi_k (t)$ [right panel] are plotted as a function of time
for $k_{\parallel} = 0$ (red), $k_{\parallel} = 1$ (blue), $k_{\parallel} = 2$ (green), and $k_{\parallel} = 5$ (magenta).} \label{inv-F2}
\end{figure}
\begin{figure}[t]
{\includegraphics[width=0.475\linewidth]{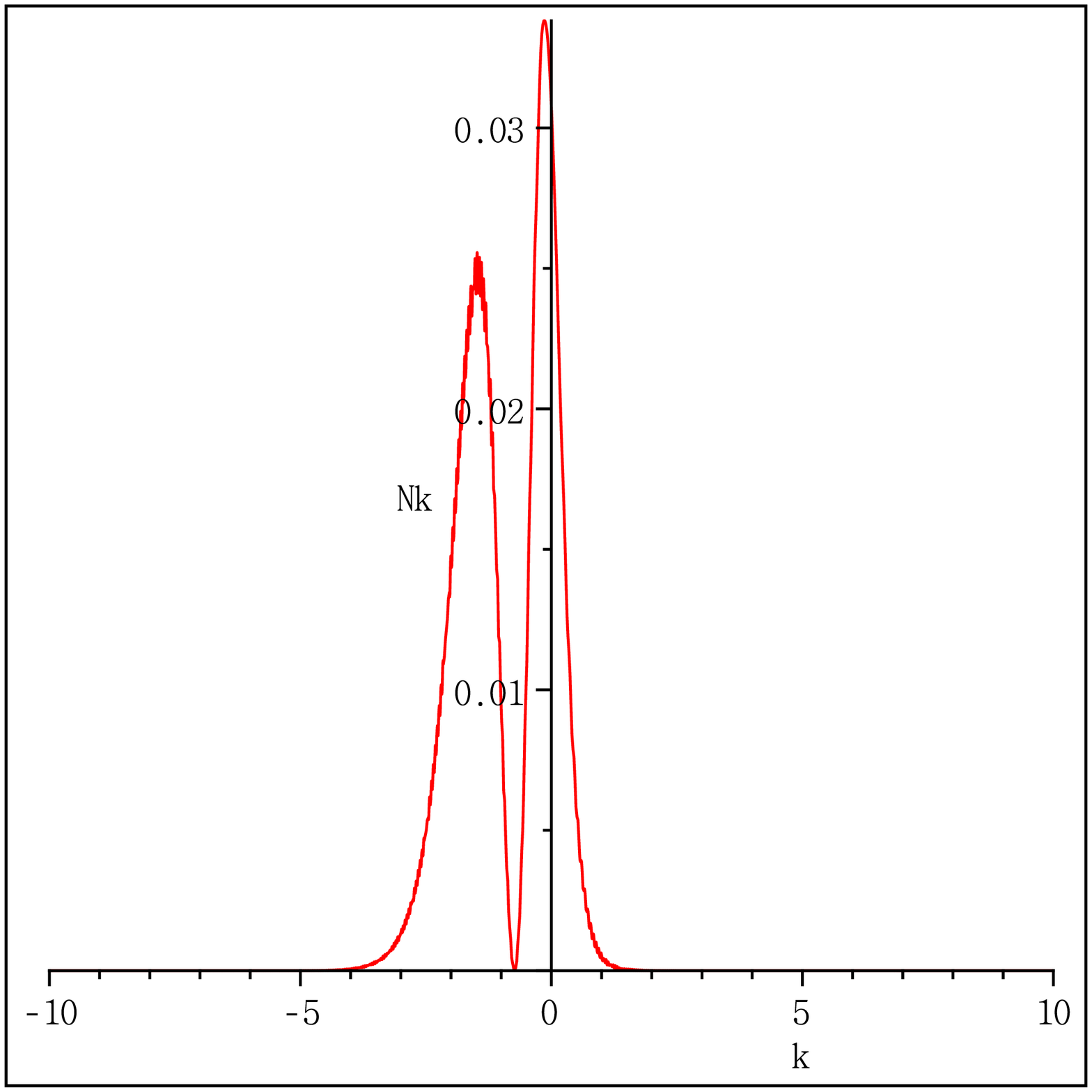}}\hfill
{\includegraphics[width=0.475\linewidth]{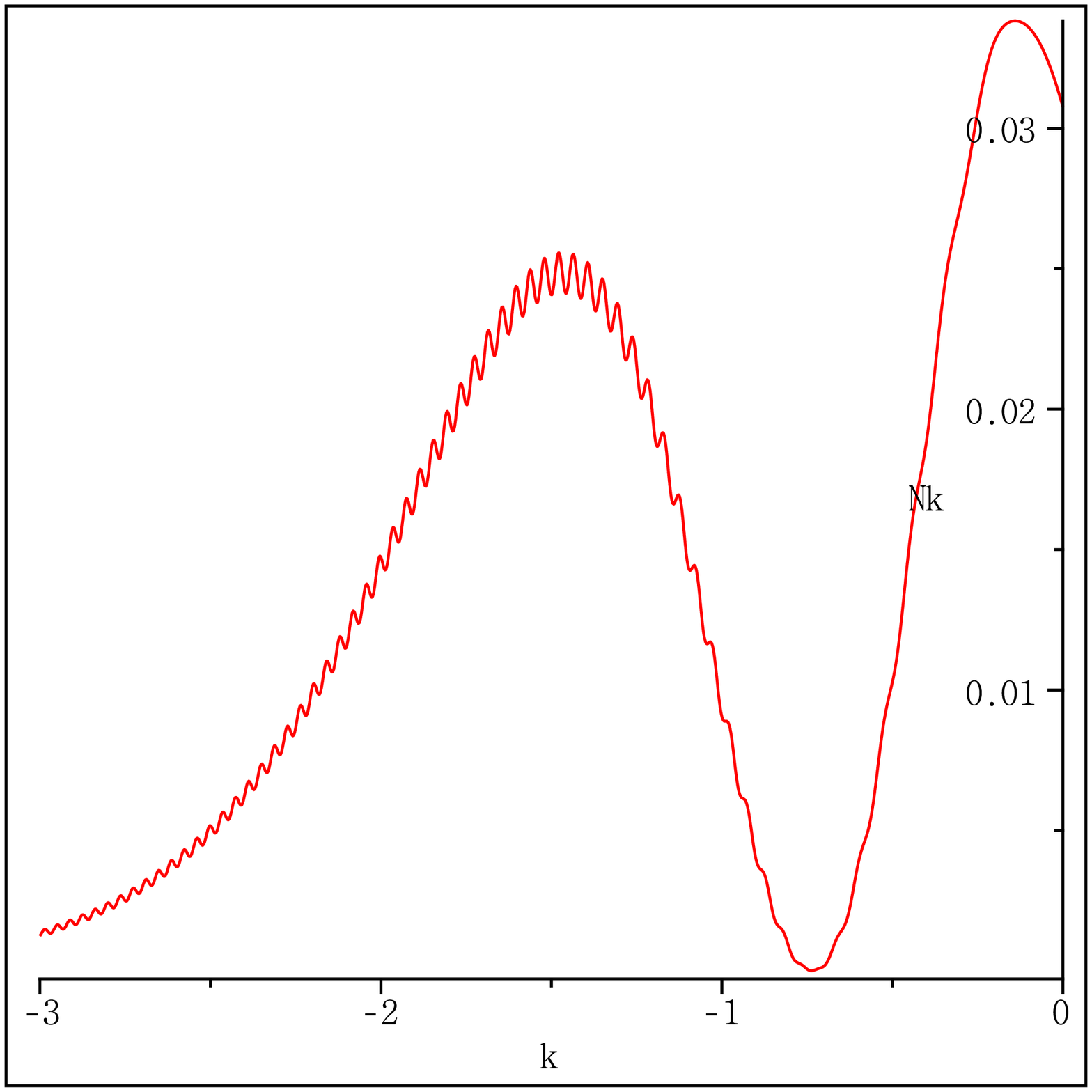}}
\caption{(color online). For the inverse square potential,  the longitudinal momentum spectrum of pairs at $t = 100$ is plotted
as a function of parallel momentum in the range of $-10 \leq k_{\parallel} \leq 10$ [left panel]
and the detailed plot $-3 \leq k_{\parallel} \leq 0$ [right panel].} \label{inv-F3}
\end{figure}

The inverse square potential shares the same property of vanishing in the remote past and future and
changing polarity as the di-polarity Sauter electric field in Sec.\,\ref{di-Saut}. The
structure of the longitudinal momentum spectrum of pairs in the inverse square potential
was discovered and explained as a consequence of Stokes phenomenon by Dumlu and Dunne \cite{DumluDunne10,DumluDunne11}.

For the numerical work we set $E_0 = 1$ and $\tau =1$. The peak of the number of pairs in Fig.\,\ref{inv-F1}
is relatively small compared with the Sauter or Gaussian electric field almost with the same peak intensity.
Also it is small
even compared with the di-polarity Sauter electric field. After the completion of interaction the number of pairs
oscillates with small amplitude as shown in Fig.\,\ref{inv-F2}. The structure of the longitudinal momentum
spectrum is shown in Fig.\,\ref{inv-F3}. The relatively simple looking spectrum has also a fine substructure as
shown in the right panel of Fig.\,\ref{inv-F3}. The result of this section cannot be directly compared with
that in Ref. \cite{DumluDunne10,DumluDunne11}, in which the structure was found for a subcritical strength and
longer time scale than ours in the Compton scale.

\subsection{$A_{\parallel} (t) = \frac{\sqrt{E_0(E_0+1)} \tau}{\cosh(\frac{t}{\tau})}$} \label{soliton}

The solitonic gauge potential has a special energy condition
\begin{eqnarray}
\omega^2 (t) = \omega^2 (- \infty) + \frac{n(n+1) \omega^2 (- \infty)}{\cosh^2(\omega (- \infty) t)},
\label{sol en}
\end{eqnarray}
where $n$ is a natural number and $\omega (- \infty)$ is the Minkowski energy of particle in the remote past.
The one-soliton $(n = 1)$ and multi-soliton $(n = 2, \cdots)$ gauge fields correspond to soliton solutions for the inverse
scattering of the Korteweg-de Vries (KdV) equation \cite{KimSchubert11,Kim11}. The solitonic gauge potential is interesting
in understanding the peculiarity of Schwinger mechanism by pulsed electric fields. Among
the gauge potentials that vanish in the remote past and future,
the solitonic gauge potential is unique in that the number of pairs produced by the corresponding electric field
exponentially increases from and then decreases to zero \cite{KimSchubert11,Kim11}. In fact,
any gauge potential which has the reflectionless scattering for the field equation (\ref{qm eq})
has zero number of pairs in the remote future and may belong to the solitonic gauge potential.
\begin{figure}[t]
{\includegraphics[width=0.475\linewidth]{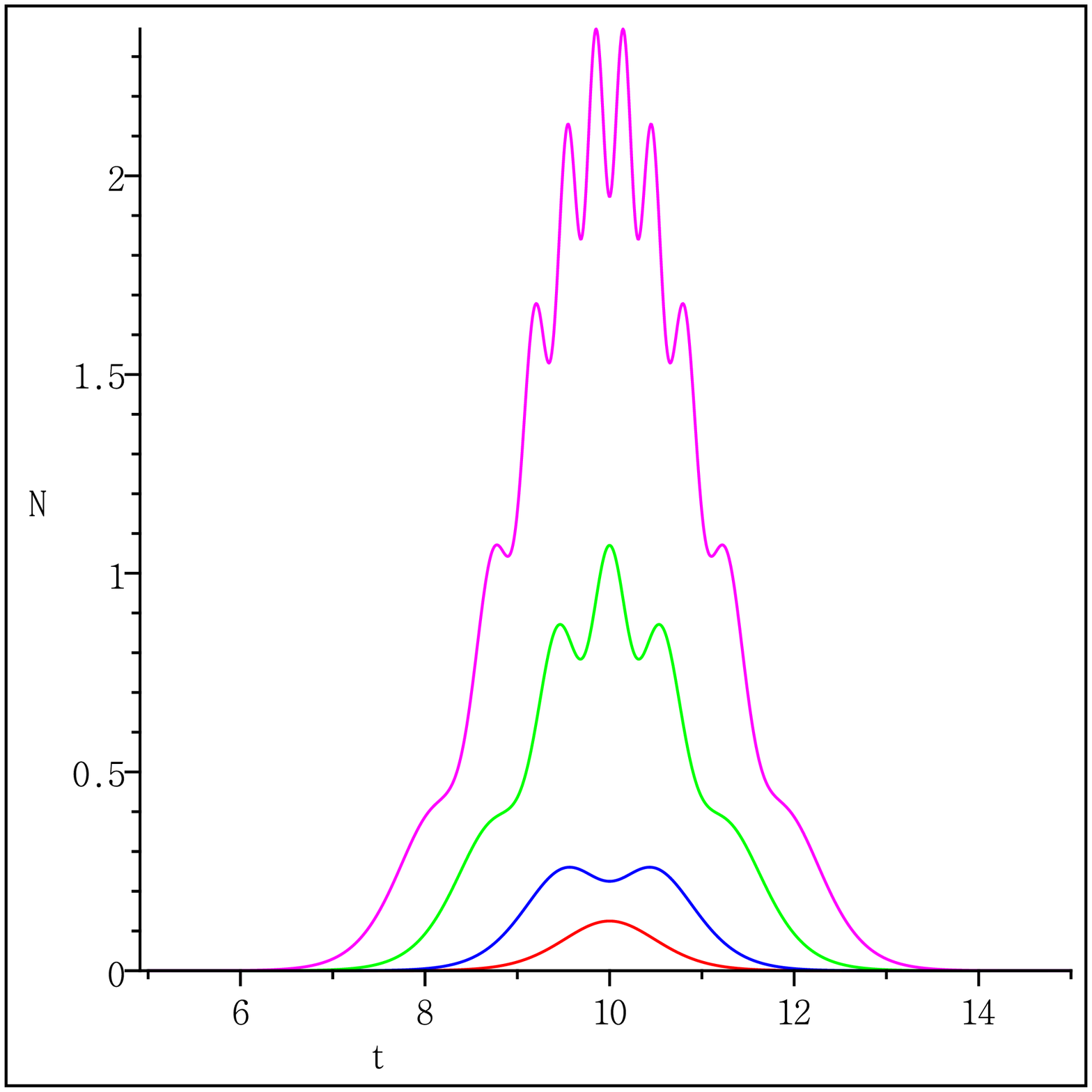}}\hfill
{\includegraphics[width=0.475\linewidth]{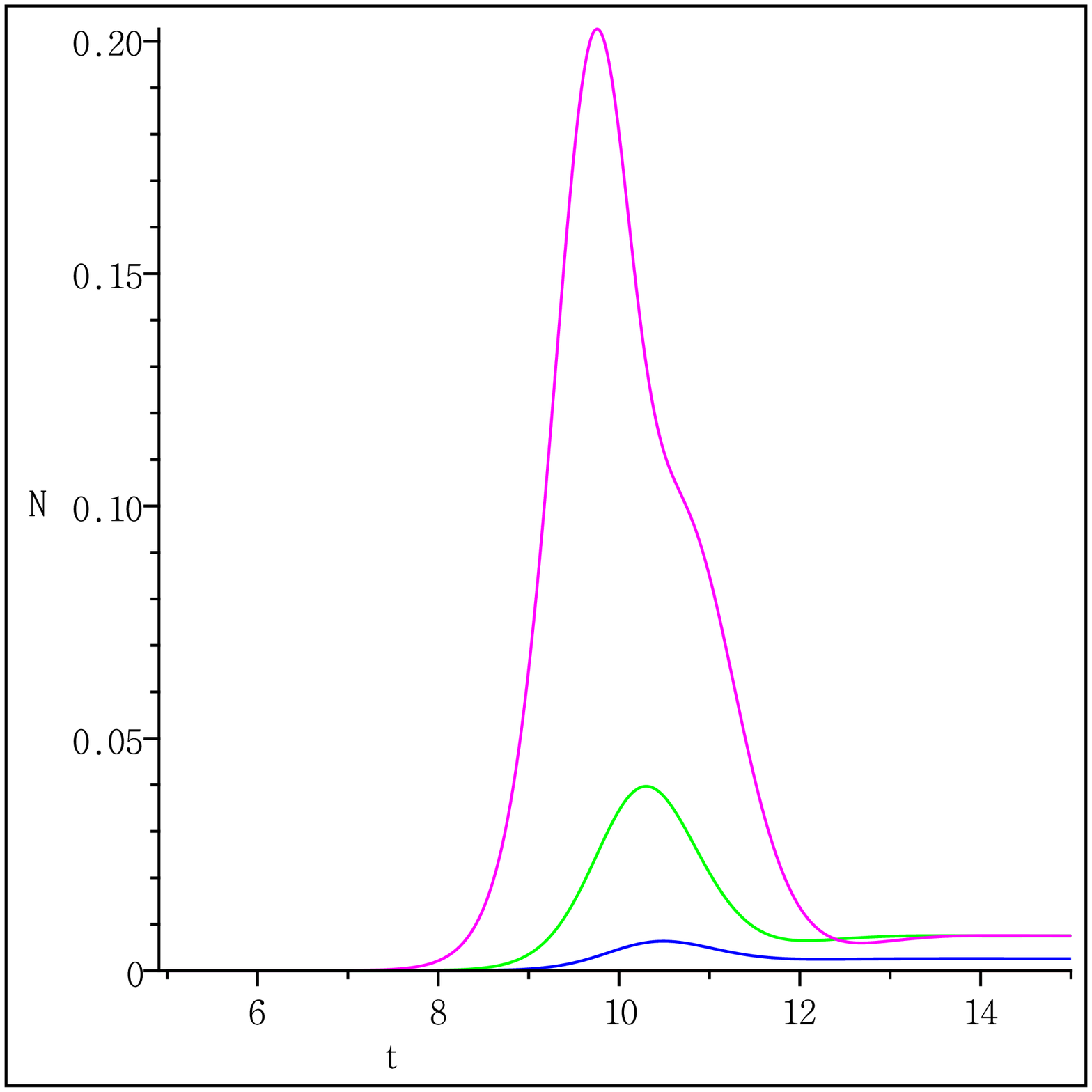}}
\caption{(color online). The number of pairs $N (t)$ is plotted for
for the solitonic gauge potential with $E_0 = 1, 2, 5, 10$ [left panel] and
for non-solitonic gauge potential with $E_0 =0.2, 0.5, 1.5$ [right panel].} \label{soliton-F}
\end{figure}

For the numerical purpose, we consider the electric field with the center shifted
\begin{eqnarray}
E(t) = \frac{\sqrt{E_0(E_0+1)} \sinh (\frac{t- 10 \tau}{\tau}))}{\cosh^2(\frac{t - 10 \tau}{\tau})},
\end{eqnarray}
and set $\tau = 1$ but select a natural number and non-natural number for the strength $E_0$.
The left panel of Fig.\,\ref{soliton-F} is the number of pairs as a function of time for
the solitonic gauge potentials $E_0 =1, 2, 5$ and $10$. The number of pairs increases and then decreases
in symmetric way around the center $t = 10$ of the gauge potential and the number of local maxima is the
same as the soliton number \cite{Kim11}. On the other hand, for a non-natural number, for instance,
$E_0 = 0.2, 0.5$ and $1.5$, the number of produced pairs increases from zero but decreases to a constant value.
This temporal behavior of the number of pairs is expected from the asymptotic solutions (\ref{asym sol1})
for $\omega (- \infty) = \omega (\infty)$ regardless of $E_0$. It should be recollected that the inverse square
potential has also finite number of pairs in the remote future in Sec.\,\ref{inv pot}, in contrast to the solitonic gauge field.

\section{Discussion and Conclusion}

We have employed the evolution operator formalism to numerically study pair production in scalar QED
by pulsed electric fields. The pulsed electric field is nontrivial in that it
acts for a finite period of time and has the inhomogeneity of time.
The Hamiltonian for a spinless charged boson in time-dependent electric field is an infinite sum
of oscillators with time-dependent frequencies. After expressing the oscillator Hamiltonian in terms of
the creation and annihilation operators of particle and antiparticle in the Minkowski vacuum, the evolution
operator (\ref{ev op}) in $SU(1,1)$ algebra is factorized into the pair annihilation part, the number part and
the pair creation part.
The advantage of this factorization is that the exact quantum state (\ref{sq vac}) from the Minkowski vacuum
under the influence of a pulsed electric field is the squeezed vacuum of particle and antiparticle
with a complex phase factor. The sum of all complex time-dependent phase factors for the number
operator determines the vacuum polarization and the vacuum persistence, which in turn is related to
the number of pairs produced by the pulsed electric field.

Now the time-dependent Schr\"{o}dinger equation is equivalent to a set of first order differential equations for
the complex parameters for the evolution operator. The evolution of the Minkowski vacuum is
governed by three parameters in Eqs.\,(\ref{gamma-I})-(\ref{xi-I}), and thus the set of first order differential equations
can implement numerical works for pair production in various configurations of the electric fields.
Remarkably the differential equations have the asymptotic solutions (\ref{asym sol1}) when
the gauge potential vanishes and another solutions (\ref{asym sol2}) when the gauge
potential approaches a constant value. These asymptotic solutions put a strong constraint
on the behavior of number of pairs such that
the number of pairs per unit volume and per unit time is a constant for the zero-gauge potential
in the remote future but it oscillates around the time-averaged value for a non-zero gauge potential.

For numerical works we have selected two classes of electric fields or gauge potentials. In the first class
the electric field does not change the polarity of the field. Therefore, the gauge potential, as
the negative of the time integral of electric field, should have a non-zero value in the remote
future. We have considered (i) Sauter electric field, (ii) Gaussian electric field and (iii) oscillating
Gaussian gauge potential as the first class of mono-polarity electric field.
In the second class the electric field changes the polarity
and has the gauge potential which vanishes in the remote future. We have considered in the second class (iv)
di-polarity Sauter electric field, two Sauter fields separated by a time gap, (v) inverse square
gauge potential and (vi) solitonic gauge potential, all of which have the di-polarity of field.

We have computed the number of pairs as a function of time and longitudinal momentum for
the pulsed electric fields of (i)-(vi) with the Compton scale.
In the first class (i)-(iii), the number of pairs increases when the field acts
and then oscillates around the time-averaged value as predicted by the asymptotic solutions.
In general the number of pairs is suppressed for large longitudinal momentum as expected from the field equation
in Sec.\,\ref{scat pic}. The structure of longitudinal momentum spectrum is found
for small momentum for (i)-(iii). Hebenstreit et al found the substructure of the spectrum
in a sinusoidal electric field with Gaussian envelope \cite{HADG09} while the sinusoidal
gauge potential is considered in this paper.
Also, the structure of the momentum spectrum has been found
for the class (v) by Dumlu and Dunne \cite{DumluDunne10,DumluDunne11}.
The number of pairs as a function time or a function of momentum
shows similarity between the Sauter electric field and the Gaussian electric field.
However, the momentum spectrum for the oscillating Gaussian electric field
reorganizes and bunches around one positive and one negative momenta with the same magnitude
and the separation in the momentum space equals
to the angular frequency of the oscillating electric field. We do not have a simple physical
explanation for this bunching effect due to the oscillating field with a Gaussian envelope.

In the second class of di-polarity field (iv)-(vi) in which the gauge potential vanishes in the remote future,
the number of pairs increases and then decreases to a constant rate and
has relatively simpler structure of the longitudinal momentum than mono-polarity electric fields.
Still the longitudinal momentum spectrum for the di-polarity Sauter electric field exhibits
a surprising feature of negative or positive momentum dominance of produced pairs
depending on whether the positively directed pulse or the negatively directed
pulse acts first, which is then followed by the oppositely directed second pulse.
Further the solitonic gauge potential produces pairs in a symmetric way in time
and returns back to the Minkowski vacuum without any residual pairs.

In this paper we have confined our study  to scalar QED in pulsed electric fields.
However, there are a few issues related to but not treated in this paper.
First, the evolution operator formalism can also apply to
spinor QED, since the spin-1/2 fermions have $SU(2)$ algebra isomorphic to $SU(1,1)$
and may have an evolution operator similar to the spinless boson.
Another issue is the vacuum polarization, the real part of
the in- and the out-vacua scattering amplitude. To get a finite effective action
one should properly regularize the sum over the momentum of the real part of the scattering
amplitude. Still another issue is the back reaction of produced pairs. The strong field investigated
in this paper produces roughly one pair per unit Compton volume, which could generate an induced electric field screening
the external field and cause another mechanism for the oscillation of produced pairs \cite{KESCM91,KESCM92}.
Finally, it would be extremely useful and promising for ELI experimentation to have a similar evolution operator
formalism for both spatially and temporally localized electric fields.
All these issues go beyond the scope of this paper and will be addressed in future publication.

\acknowledgments

S.~P.~K. and H.~W.~L. would like to thank Professor Remo Ruffini for the warm hospitality at
ICRANet where this paper was completed.
The work of S.~P.~K. was supported in part by Basic Science Research Program through
the National Research Foundation of Korea (NRF) funded by the Ministry of Education, Science and
Technology (22012R1A1B3002852). The work of H.~W.~L was supported in part by the International Research and Development Program of the
National Research Foundation of Korea (NRF) funded by the Ministry of Education, Science
and Technology (MEST) of Korea(K21003002081-12B1200-00310). The work of R.~R. was supported in part by ICRANet.
The visit to ICRANet by S.~P.~K. and H.~W.~L. was supported in part by ICRA.

\end{document}